\documentclass{LMCS}

\def\dOi{11(3:14)2015}
\lmcsheading%
{\dOi}
{1--32}
{}
{}
{Oct.~\phantom02, 2014}
{Sep.~17, 2015}
{}

\ACMCCS{[{\bf Theory of computation}]: Formal languages and automata theory}

\usepackage{hyperref}
\usepackage{xspace}
\usepackage{amssymb,amsmath,stmaryrd}
\usepackage{amsfonts}
\usepackage{amssymb}
\usepackage{wasysym}
\usepackage{bbold}
\usepackage[lined,boxed,noend,ruled,resetcount,norelsize]{algorithm2e}

\usepackage{xcolor,pgf,tikz,pgflibraryarrows,pgffor,pgflibrarysnakes}
\usepackage{booktabs}

\usetikzlibrary{automata}
\usetikzlibrary{snakes,backgrounds,positioning,shadows}
\usetikzlibrary{patterns}
\usetikzlibrary{shapes}

\newcommand{\Sum}{{\sf Sum}\xspace}
\newcommand{\Dsum}{{\sf Dsum}\xspace}
\newcommand{\Avg}{{\sf Avg}\xspace}
\newcommand{\Ratio}{{\sf Ratio}\xspace}

\newcommand{\ptime}{{\sf PTime}\xspace}
\newcommand{\del}{{\sf delay}}
\newcommand{\delay}{\del}
\newcommand{\dom}{\text{dom}}

\newcommand{\run}[1]{\xrightarrow{#1}}

\def\abs#1{\ensuremath{\lvert #1\rvert}}
\newcommand{\nat}{\mathbb{N}}
\newcommand{\rat}{\mathbb{Q}}

\newcommand{\outcome}{\mathsf{outcome}}
\newcommand{\minusone}{\scalebox{1.2}[1]{-}1}

\begin{document}

\title[Quantitative Languages Defined by Functional
  Automata]{Quantitative Languages Defined by Functional
  Automata\rsuper*}

\author[E.~Filiot]{Emmanuel Filiot\rsuper a}	
\address{{\lsuper{a,c}}Universit\'e Libre de Bruxelles, Computer Science Department}	
\email{\{efilio, jraskin\}t@ulb.ac.be}  
\thanks{{\lsuper a}Emmanuel Filiot is supported by F.R.S.-FNRS (as a research associate, ``Chercheur Qualifi\'e'')}	

\author[R.~Gentilini]{Raffaella Gentilini\rsuper b}	
\address{{\lsuper b}University of Perugia, Department of Mathematics and Computer
Science}	
\email{raffaella.gentilini@dmi.unipg.it}  
\thanks{{\lsuper b}Raffaella Gentilini acknowledges support from the Italian National Group for Scientific Computing (GNCS-INdAM)}	

\author[J.-F.~Raskin]{Jean-Fran\c{c}ois Raskin\rsuper c}	
\address{\vspace{-18 pt}}	
\thanks{{\lsuper c}Jean-Fran\c{c}ois Raskin is supported by ERC Starting Grant (279499:
    inVEST)}	

\titlecomment{{\lsuper*}This work is a revised and  extended version of
    \cite{FGR12}}

\begin{abstract}
A weighted automaton is functional if any two accepting runs on the same finite word 
have the same value. In this paper, we investigate functional weighted automata for four
different measures: the sum, the mean, the discounted sum of weights along
edges and the ratio between rewards and costs. On the positive side, we
show that functionality is decidable for the four measures. Furthermore,
the existential and universal threshold problems, the language
inclusion problem and the equivalence problem  are all decidable when the weighted automata are functional.
On the negative side, we also study the quantitative
extension of the realizability problem and show that it is undecidable for
sum, mean and ratio. We finally show how to decide whether the language associated with a given
functional automaton can be defined with a deterministic one, for sum, mean and discounted sum.
The results on functionality and determinizability are expressed for the more general class of
functional group automata. This allows one to formulate within the same framework new results related
to discounted sum automata and known results on sum and mean automata. 
Ratio automata do not fit within this general scheme and different
techniques are required to decide functionality.
\end{abstract}

\maketitle

\section{Introduction}
Recently, there have been several efforts made to lift the foundations
of computer aided verification and synthesis from the basic {\em
Boolean} case to the richer {\em quantitative} case,
e.g.~\cite{ChatterjeeDH10,BokerCHK11,AminofKL11}. This paper belongs
to this line of research and contributes to the study of quantitative
languages over finite words.

Our paper proposes a systematic study of the algorithmic properties of
several classes of {\em functional} weighted automata (defining
quantitative languages). A functional weighted automaton is a {\em
nondeterministic} weighted automaton such that any two accepting runs $\rho_1$,
$\rho_2$ on a word $w$ associate with this word a unique value ${\sf
V}(\rho_1)={\sf V}(\rho_2)$. As we show in this paper, several
important verification problems are decidable for nondeterministic
functional weighted automata while they are undecidable (or not known
to be decidable) for the full class of nondeterministic weighted
automata. 
\enlargethispage{\baselineskip}

In this paper, we study automata in which an integer weight, or a pair
of integer weights, is associated with each of their transitions. From
those weights, an (accepting) run $\rho$ on a word $w$ associates a
sequence of weights with the word, and this sequence is mapped to a
rational value by a {\em measure function}. We consider four different
measure functions\footnote{We do not consider the measure functions
{\sf Min} and {\sf Max} that map a sequence to the minimal and the
maximal value that appear in the sequence as the nondeterministic
automata that use those measure func\-tions can be made deterministic
and all the decision problems for them have known and simple
solutions \cite{ChatterjeeDH10}.}: $(i)$ \Sum computes the sum of the weights along the
sequence, $(ii)$ \Avg returns the mean value of the weights,
$(iii)$ \Dsum computes the discounted sum of the weights for a given
discount factor $\lambda \in \rat\cap ]0,1[$, and $(iv)$ \Ratio is applied to a
sequence of pairs of weights, and it returns the ratio between the sum
of weights appearing as the first component (rewards) and the sum of
the weights appearing as the second component (costs). All those measures are motivated by
applications in computer aided verification and synthesis, see for
example~\cite{AlfaroFHMS05,BloemGHJ09}. The value
associated with a word $w$ is obtained by combining all the values 
of the accepting runs on $w$ with a particular operation (usually max or
min). The value of $w$ is denoted by $L_A(w)$. 

\vspace{2mm}
\noindent \textbf{Contributions}
Classical results on weighted automata consider operations
over semirings: the value of a run is obtained as the multiplication
of the values along its transitions, and the values of all runs on the same
input word are combined with addition \cite{Droste_Kuich_Vogler_2009}. Since we focus on functional
automata, all the accepting runs have the same value, and so we do not need
addition. Whenever it is possible, we phrase our results in the
general framework of functional group automata, i.e. automata whose
transitions are weighted by elements of a group. In
particular, \Sum, \Avg, and \Dsum can be seen as operations over a
group. For \Ratio however, we always need different techniques, and
leave its encoding in terms of a group operation as open.

We first show that functionality is decidable in \textsf{PTime} for
group automata (operations on group elements are
assumed to be computable in polynomial time). This implies that functionality is
\textsf{PTime} for \Dsum automata and generalizes known results for \Sum
and \Avg automata. 
By using a pumping argument, we show that functionality is
in \textsf{CoNP} for \Ratio-automata.

Then we solve the following decision problems, along the line of~\cite{ChatterjeeDH10}. First, we
consider {\em threshold} problems. The {\em existential} ({\em
universal}, respectively) {\em threshold} problem asks, given a
weighted automaton $A$ and a threshold $\nu \in \rat$, if there exists
a word (if for all words, respectively) $w$ accepted by $A$:
$L_A(w)  \geq \nu$. Those problems can be seen as generalizations of the
emptiness and universality problems for finite-state automata. Second,
we consider the {\em quantitative language inclusion problem} that
asks, given two weighted automata $A$ and $B$, if all words accepted
by $A$ are also accepted by $B$, and for all accepted words $w$ of
$A$, we have $L_A(w) \leq L_B(w)$. We show that all those problems are
decidable for the four classes of measure functions that we consider
in this paper when the automata are functional. In particular, 
for \Ratio, we show decidability of the inclusion problem using a
recent algorithm to solve quadratic diophantine equations~\cite{Grunewald04},
this is a new deep result in mathematics and the complexity of the
algorithm is not yet known. We also show that the equivalence problem
can be decided in polynomial space for \Ratio via an easy reduction to
functionality. Note that those decidability results are
in sharp contrast with the corresponding results for the full class of
nondeterministic weighted automata: for that class, only the
existential threshold problem is known to be decidable, the language
inclusion problem is undecidable for \Sum \cite{Krob/94,AlmagorBokerKupferman2011}, and therefore for
\Avg and \Ratio,  while the problem is open for \Dsum.

Finally, we consider a (finite word) quantitative variant of the {\em realizability}
 problem introduced by Church, which is 
related to the synthesis of reactive systems \cite{PnuRos:89,DBLP:conf/birthday/Thomas08} and can be
formalized as a game in which two players (the system and the environment) alternate in choosing
letters in their respective alphabet of signals. The system can decide
to stop the game. By doing so, they form a word
which is obtained by concatenating the successive choices of the
players. The realizability problem asks, given a weighted automaton
$A$, a threshold $\nu\geq 0$, and an alphabet $\Sigma=\Sigma_1 \times \Sigma_2$, if there exists a
strategy for choosing the letters in $\Sigma_1$ in the word forming
game such that no matter how the adversary chooses his letters in
$\Sigma_2$, the word $w$ that is obtained belongs to the language of
$A$ and $A(w) > \nu$. We show that this problem is undecidable for
$\Sum$, $\Avg$, and $\Ratio$ functional automata (the case \Dsum is left open). However, we show that the realizability
problem is decidable for the deterministic versions of the
automata studied in this paper. This motivates the {\em
determinizability} problem.

 The determinizability problem asks, given a functional
weighted automaton $A$, if the quantitative language defined by $A$ is
also definable by a {\em deterministic} automaton. It is known
that \Sum, \Avg and \Dsum-automata are not determinizable in
general \cite{ChatterjeeDH10}. We give here a decidable
{\em necessary} and {\em sufficient} condition for the
determinizability of  functional group automata, and we
show how to construct a deterministic automaton from the functional
one when this is possible. As a corollary, we obtain a decidable 
characterization of determinizable functional \Sum-, \Avg- and \Dsum-automata.
While it was known for \Sum (and as a consequence for \Avg, by seing the
\Avg-automaton as a \Sum-automaton) \cite{KirstenM05}, it is new for \Dsum.

\vspace{2mm}
\noindent \textbf{Functionality versus Unambiguity} 
Functional weighted automata are a natural generalization
of {\em unambiguous} weighted automata, i.e. weighted automata such
that there is at most one accepting run for each input word. Since unambiguity captures
most of the nondeterminism that is useful in practice, our results are
both theoretically and practically important. Functional weighted
automata are equivalent (modulo an exponential blow-up) to unambiguous
weighted automata. This exponential blow-up is worst-case
unavoidable. It is already unavoidable when going from a
non-deterministic to an unambiguous finite automaton
\cite{Schmidt77}. Therefore, we inherit this succinctness result.

Having algorithms to test for functionality has the nice consequence
of providing algorithms to test for equivalence of functional (and
hence unambiguous) weighted automata: given two functional
weighted automata $A_1$ and $A_2$, they are equivalent iff 
they have the same domain, and their union is functional. For
\Ratio-automata for instance, this gives better complexity than
applying our result on inclusion (which is, in some sense, a harder
problem). 

In all the decision problems we consider but functionality,
considering unambiguous weighted automata instead of functional ones
would not have simplified the algorithms nor the proofs, because in a
way, unambiguous automata capture already all the difficulty of
non-deterministic behaviours, and are, therefore, as difficult as
functional automata to deal with. Therefore, considering functional
automata with respect to decision problems comes, in some sense, for
free.


\noindent \textbf{Related Works} 
Motivated by open problems in computer-aided verification, our work
follows the same line as \cite{ChatterjeeDH10}. However
\cite{ChatterjeeDH10} is concerned with weighted automata on
infinite words, either non-deterministic, for which some important problems
are undecidable (e.g. inclusion of \Avg-automata), or deterministic ones, which
are strictly less expressive than functional automata. The \Ratio
measure is not considered either. Their domains of
quantitative languages are assumed to be total (as all states are
accepting and their transition relation is total) while we can 
define partial quantitative languages thanks to an acceptance
condition. Functional weighted automata where all states are accepting
correspond to deterministic weighted automata: all the transitions that leave the same state on
the same input symbol must carry the same weight, and therefore, a
simple subset construction allows one to transform any such functional
automaton into a deterministic one.

Except for realizability, our results for \Sum-automata 
(and to some extent \Avg-auto\-mata) are not new \cite{Mohri09}. Functionality is known to be
in {\sf PTime} \cite{KirstenM05}, and emptiness, inclusion,
equivalence (for functional \Sum-automata) are already known to be
decidable \cite{Krob94someconsequences,DBLP:journals/ita/LombardyM06}. Moreover, it is known
that determinizability of functional \Sum-automata is decidable in {\sf
PTime} \cite{KirstenM05}, as well as for the strictly more expressive
class of polynomially ambiguous \Sum-automata \cite{journals/mst/Kirsten06}, for which 
the termination of Mohri's determinization algorithm \cite{Droste_Kuich_Vogler_2009} is
decidable. Weighted automata over semirings have been extensively studied~\cite{Droste_Kuich_Vogler_2009}, and more generally rational
series \cite{BerstelReutenauer88}. Mohri's determinization algorithm
has been generalized in \cite{KirstenM05} to arbitrary semirings, in which a general
condition for its termination, generalizing the notion of twins property in \cite{Choffrut77,Mohri97}, is given. However, this sufficient
condition only applies to commutative semirings, and therefore cannot directly
be used for \Dsum-automata. 
We rephrase the twinning property on groups that are not necessarily
commutative and prove that it is a
sufficient  condition for a functional group automata
to be determinizable. Our
determinization algorithm for functional group automata is similar to
Mohri's algorithm and is, in that sense, not new. However, since we
cannot rely on the additive operation, we need some different
computation of the output values of the constructed deterministic
automaton. Further, we extend \cite{KirstenM05} from functional \Sum-automata to functional \Avg and \Dsum-automata, proving that the twinning property is also a necessary condition toward determinization for these measures.

\noindent The techniques we use for deciding functionality and determinization are
also inspired by techniques from word
transducers
\cite{Schutz75,JCSS::BlattnerH1977,BealEtAl03a,Choffrut77,DBLP:journals/iandc/WeberK95}. 
In particular, our procedure to decide functionality of weighted
automata also allows us to decide functionality of a word transducer, 
seen as a weighted automaton over the free group. 
It generalizes to arbitrary groups the procedure of \cite{BealEtAl03a}
which was used to show that functionality of word transducers is in
{\sf PTime}. As in \cite{BealEtAl03a}, it relies on a notion of delay
between two runs. This notion of delay is also used for the
determinization of group automata.

In \cite{DBLP:conf/csl/BokerH11}, Boker
et. al. show that \Dsum-automata on infinite words with a trivial
accepting condition (all states are accepting), but not necessarily
functional, are determinizable for any discount factor of the form $1/n$ for some
$n\in\mathbb{N}_{\geq 2}$, while we consider arbitrary discounted factors. Their proof is based on a notion
of \textit{recoverable gap}, similar to that of delays. 
Finally in \cite{DBLP:conf/wia/DrosteR07}, the relation between discounted 
weighted automata over a semiring and weighted logics is studied.

To the best of our knowledge, our results on \Dsum and \Ratio-automata, as well
as on the realizability problem, are new. Our main and most technical results are
functionality and inclusion of \Ratio-automata, undecidability of the
realizability of functional \Sum-automata, and solvability of the deterministic versions of the realizability
problem. The latter reduce to games on graphs that are to the best of our
knowledge new, namely finite $\Sum,\Avg,\Dsum,\Ratio$-games on weighted
graphs with a combination of a reachability objective and a
quantitative objective.

\section{Quantitative Languages and Functionality} 
\newcommand{\talpha}{\Sigma_\dashv^+}

Let $\Sigma$ be a finite alphabet. We denote by $\Sigma^*$ the set of
finite words over $\Sigma$, and $\Sigma^+$ the set of non-empty words
over $\Sigma$. The empty word is denoted by $\epsilon$. In this paper,
we assume that weighted automata process input words that end with a
terminal symbol ${\dashv}\in\Sigma$. We denote by $\talpha$ the
set of words of the form $u{\dashv}$, where $u\in(\Sigma\setminus
\{{\dashv}\})^*$. 

A \textit{quantitative language L} over 
$\Sigma$ is a mapping $L: \talpha\rightarrow \mathbb{Q}\cup
\{\perp\}$\footnote{As in \cite{ChatterjeeDH10}, we do not consider the empty word as our weighted
automata do not have initial and final weight functions. This eases
our presentation but all our results carry over to the more
general setting with initial and final weight
function \cite{Droste_Kuich_Vogler_2009}.}. 
For all $w\in\talpha$, $L(w)$ is called the
\textit{value} of $w$. $L(w) = \perp$ means that
the value of $w$ is undefined. We set $\perp < v$ for all $v\in \mathbb{Q}$.


Let $n\geq 0$. Given a finite sequence $v = v_0\dots v_n$ of integers (resp. a finite
sequence $v' = (r_0,c_0)\dots(r_n,c_n)$ of pairs of natural numbers,
$c_i>0$ for all $i$) and $\lambda\in\mathbb{Q}$ such
that $0<\lambda< 1$, we define the following
functions:
\vspace{-2mm}
$$
\Sum(v)= \sum_{i=0}^n v_i \quad \Avg(v)= \frac{\Sum(v)}{n+1} \quad
\Dsum(v)= \sum_{i=0}^n\lambda^iv_i \quad 
 \Ratio(v')= \dfrac{\sum_{i=0}^n r_i}{\sum_{i=0}^n c_i}
$$
\vspace{-2mm}


\noindent \textbf{Weighted Automata} Let
$V{\in}\{\Sum,\Avg,\Dsum,\Ratio\}$. A \textit{weighted $V$-automaton} over
$\Sigma$ is a tuple $A {=} (Q,q_I,F,\delta,\gamma)$ where $Q$ is a
finite set of states, $q_I$ is an initial state, $F$ is a set of
final states, $\delta =  \delta'\cup \delta_\dashv$ is a transition
relation, where $\delta'\subseteq (Q\setminus F){\times}
(\Sigma\setminus \{\dashv\}) {\times} (Q\setminus F)$ and 
$\delta_\dashv\subseteq (Q\setminus F){\times}
\{\dashv\} {\times} F$ is the terminal set of transitions, and
$\gamma:\delta \rightarrow \mathbb{Z}$
(resp. $\gamma:\delta\rightarrow \mathbb{N}{\times} (\mathbb{N}-0)$ if
$V=\Ratio$) is a \textit{weight function}. The size of $A$ is defined by $|A| = |Q|+|\delta|
+ \sum_{t\in\delta} log_2(\gamma(t))$. Note that $(Q,q_I,F,\delta)$ is
a classical finite-state automaton, called the input-automaton. 
We say that $A$ is \textit{deterministic} if its input-automaton
$(Q,q_I,F,\delta)$ is deterministic.  In the sequel, we use the term $V$-automata to denote either $\Sum$, $\Dsum$, $\Avg$
or $\Ratio$-automata.

A run $\rho$ of $A$ over 
$w=\sigma_0\dots\sigma_n\in\Sigma^+$ is a sequence
$\rho = q_0\sigma_0q_1\dots\sigma_nq_{n+1}$ such that
$q_0=q_I$ and for all $i\in\{0,\dots,n\}$,
$(q_I,\sigma_{i},q_{i+1})\in\delta$. It is \textit{accepting} if
$q_{n+1}\in F$. We write $\rho:q_0\run{w} q_{n+1}$ to denote that $\rho$ is a
run on $w$ starting at $q_0$ and ending in $q_{n+1}$. 
Given $i\in\{0,\dots,n\}$, we write  $\rho_i$
to denote the prefix of the run $\rho$
 until position $i$.
The domain of $A$, denoted by $\dom(A)$, is defined
as the set of words $w\in\Sigma^*$ on which there exists some accepting run of
$A$. Note that by definition of $\delta$, we have $\dom(A)\subseteq
\talpha = (\Sigma\setminus \{ \dashv\})^*{\dashv}$. We say that $A$ is \textit{unambiguous}  if
it admits at most one accepting run for each word.

The function $V$ is naturally extended to runs as follows:
$$
V(\rho)\ =\ \left\{\begin{array}{llllll}
     V(\gamma(q_0,\sigma_0,q_1)\dots \gamma(q_{n},\sigma_n,q_{n+1})) &
     \text{ if $\rho$ is accepting} \\
     \perp & \text{ otherwise}
\end{array}\right.
$$

The  relation  $R_A^V\
  =\ \{(w,V(\rho))\ |\ w\in\Sigma^+, \rho \text{ is an accepting run of
  $A$ on $w$}\}$ is called the \textit{relation induced by $A$}.
It is \textit{functional} if  for all words $w\in\talpha$, we have
$|\{v\ |\ (w,v)\in R_A^V\}|\leq 1$. In that case we say
  that $A$ is functional. The \textit{quantitative language} $L_A:\talpha\rightarrow
\mathbb{Q}\cup \{\perp\}$ defined by $A$ is defined by
$L_A: w\mapsto \max\{ v\ |\ (w,v)\in R_A^V\}$ where $\max\ \emptyset =
\perp$.

\begin{figure}[t]
\centering
\vspace{-9mm}

\begin{tikzpicture}[->,>=stealth',shorten >=1pt,auto,node distance=2.6cm,
                    semithick,scale=0.8]
  \tikzstyle{every state}=[fill=gray!20!white]

  \node[initial,initial by arrow, initial text={},initial above, state] at (-2.5,0) (qI) {$q_I$} ;
  \node[state] at (-5,0) (qa)  {$q_a$} ;
  \node[state] at (0,0) (qb) {$q_b$} ;
  \node[accepting,state] at (-2.5,-2) (qf) {$q_f$} ;

  \path (qI) edge [above] node {$a|1,b|0$} (qa) ;
  \path (qI) edge [above] node {$a|0,b|1$} (qb) ;
  \path (qa) edge [loop above] node {$a|1,b|0$} (qa) ;
  \path (qb) edge [loop above] node {$a|0,b|1$} (qb) ;

  \path (qa) edge [left] node {${\dashv}|0$} (qf) ;
  \path (qb) edge [right] node {${\dashv}|0$} (qf) ;
  \path (qI) edge [left] node {${\dashv}|0$} (qf) ;

  \node[initial,initial text={},initial above,state] at (6,0) (pI) {$p_I$} ;
  \node[state] at (8.5,0) (p) {$p$} ;
  \node[state] at (3.5,0) (q) {$q$} ;
  \node[state] at (8.5,-2) (paa) {$p_a$} ;
  \node[state] at (3.5,-2) (qbb) {$q_b$} ;
  \node[accepting,state] at (6,-2) (qff) {$q_f$} ;

  \path (pI) edge [above] node {$a|1,b|0$} (p) ;
  \path (pI) edge [above] node {$a|0,b|1$} (q) ;
  \path (p) edge [loop above] node {$a|1,b|0$} (p) ;
  \path (q) edge [loop above] node {$a|0,b|1$} (p) ;
  \path (p) edge node {$a|1$} (paa) ;
  \path (q) edge node {$b|1$} (qbb) ;
  \path (qbb) edge [below] node {${\dashv}|0$} (qff) ;
  \path (paa) edge [below] node {${\dashv}|0$} (qff) ;






\end{tikzpicture}

\vspace{-5mm}
 \caption{Examples of \Sum-automata}\label{fig:sumautomata}
\vspace{-5mm}
\end{figure}
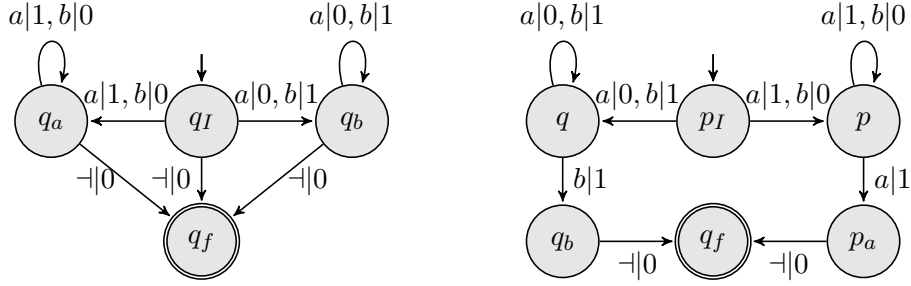

\begin{exa}
        Fig. \ref{fig:sumautomata} illustrates two \Sum-automata over
        the alphabet $\{a,b,{\dashv}\}$. The first automaton (on the left)
        defines the quantitative language
        $w\in\talpha\mapsto \max(\#_a(w),\#_b(w))$, where
        $\#_\alpha(w)$ denotes the number of occurrences of the letter
        $\alpha$ in $w$. Its induced relation is $\{ (w,\#_a(w))\mid
        w\in\talpha\}\cup \{ (w,\#_b(w))\mid
        w\in\talpha\}$. The
        second automaton (on the right) defines the quantitative
        language that maps any word of length at least 2 to the number
        of occurrences of its last letter but one. 
\end{exa}

\begin{rem}
    In the literature, weighted automata are sometimes equipped with a
    terminal function that associates with accepting states a
    value. Instead, we assume that weighted automata accepts only
    words that end with the terminal symbol $\dashv$. The two models
    are equivalent in the following sense: we can always transform a
    weighted automaton $A$ over an alphabet $\Gamma$ with a terminal
    function, into a weighted
    automaton $A_\dashv$ over the alphabet $\Gamma\cup \{\dashv\}$
    without terminal function, such that
    $\dom(A_\dashv) = \dom(A){\dashv}$ and such that all the decision
    problems we consider are preserved, such as functionality,
    emptiness, determinizability, etc. In other words, $A$ is
    functional iff $A_\dashv$ is, etc. This is done by adding
    transitions from accepting states $q_f$ of $A$ to a fresh accepting
    state of $A_\dashv$ on reading $\dashv$, with values $t(q_f)$ if
    $t$ is the terminal function. We rather use terminal symbols instead of terminal
    functions since it lightens the notations in our proofs.

    All our results but determinizability hold true for weighted
    automata without terminal function nor terminal symbol, because 
    the latter can easily be encoded as weighted automata with a
    terminal symbol, while preserving the properties we are interested
    in. For determinization however, our determinization procedure, if
    it terminates, does not necessarily produce a deterministic
    weighted automaton without terminal function/symbol, even if we
    apply it to a weighted automaton without terminal function/symbol. 
    The deterministic weighted automata that are constructed by our
    procedure coincide with the classical notion of 
    subsequential weighted automata with
    terminal function that is sometimes used in the literature
    \cite{Droste_Kuich_Vogler_2009}. Without terminal function/symbol,
    deterministic weighted automata coincide with sequential weighted
    automata. Since the notion of subsequentiality is more general
    than sequentiality, and since it already guarantees
    decidability of the realizability problem, we rather consider
    this more general class. 
\end{rem}

A state $q$ is  \textit{accessible}
 (by some
word $w\in\Sigma^*$) if $A$ admits  a run $\rho:q_I\run{w} q$. A state $q$ is  \textit{co-accessible}
(by some
word $w\in\Sigma^*$) if $A$ admits  a run $\rho:q\run{w} q_f$ for
some $q_f\in F$. 
We  say that a state $q$ is \emph{useful} if it is both accessible and co-accessible (and \emph{useless}, otherwise). Useless states  can be removed  from the given weighted automaton in linear time, without changing the recognized language \cite{Mohri09}. A weighted automaton  with no useless state is said to be \emph{trim}.

A pair of states $(q,q')$ is co-accessible if
there exists a word $w$ such that $q$ and $q'$ are co-accessible by
$w$.

\vspace{2mm}
\noindent \textbf{Functional Weighted Automata} The
\Sum-automaton on the left of Fig. \ref{fig:sumautomata} is not
functional (e.g. the word $abb{\dashv}$ maps to the values 1 and 2),
while the one on the right is functional (and even unambiguous).
      Concerning the expressiveness of functional automata, we can
      show that deterministic automata are strictly less expressive
      than functional automata which are again strictly less expressive
      than the full class of weighted automata. 
Let $V\in\{\Sum,\Avg,\Ratio\}$. The automata of
Fig. \ref{fig:sumautomata} can be seen as $V$-automata (with a
constant cost $1$ if $V = \Ratio$). The right $V$-automaton
cannot be expressed by 
            any deterministic $V$-automaton because the value of a word depends on
      its   last letter. The left $V$-automaton  cannot be expressed
      by any functional $V$-automaton. It is easy to verify that the above results
       hold also for $\Dsum$-automata.
      Therefore, for the four measures that we consider, i.e. for  $V\in\{\Sum,\Avg,\Ratio, \Dsum\}$, deterministic $V$-automata
      are strictly less expressive than functional $V$-automata which are again strictly
      less expressive than the full class of $V$-automata.

The following result shows that unambiguous $V$-automata are as
expressive as functional $V$-automata, modulo an exponential
blow-up. This blow-up is worst-case unavoidable, already 
for non-deterministic and unambiguous finite state automata
\cite{Schmidt77}. 

\begin{lem}\label{lem:functounamb}
        Let $V\in\{\Sum,\Avg,\Dsum,\Ratio\}$. For all functional
        $V$-automaton with $n$ states we can construct an equivalent
        unambiguous $V$-automaton with $O(n\cdot2^n)$ states.
\end{lem}

\begin{proof}
  Our proof is independent on the measure. Let $A =
  (Q,q_I,F,\delta,\gamma)$ be a functional $V$-automaton. We order
  the transitions of $\delta$ by a total order denoted
  by $<_\delta$. We construct an  unambiguous automaton
  $A' = (Q',q'_I,F',\delta',\gamma')$ equivalent to $A$, where:
  \begin{itemize}
    \item $Q' = Q\times 2^Q$ ;
    \item $q'_i = (q_I,\varnothing)$ ;
    \item $F' = F\times\{ P\subseteq Q\ |\ F\cap P=\varnothing\}$ ;
    \item $\gamma'\ :\ ((p,P),a,(p',P'))\mapsto \gamma(p,a,p')$ ;
  \end{itemize}

  Before defining $\delta'$ formally, let us explain intuitively the
  semantics of the states in $Q'$. The automaton $A'$ will guess a 
  run of $A$ on first state component (called the \textit{current run}).
  A pair $(p,P)$ represents the
  state $p$ of current run in the original automaton $A$ while $P$
  represents the states reached by all the runs that are greater than 
  the current run (w.r.t. the order $<_\delta$ lexigraphically extended
  to runs). At the end of the word, the run is accepting iff $p$ is
  accepting and there is no accepting state in $P$. In other words, a
  run of $A'$ on a word $w$ is accepting iff the run it defines on the first
  component is the biggest accepting run of $A$ on $w$.

  When a new letter $a\in\Sigma$ is read, $A'$ guesses a 
  transition from $p$ to some state $p'$, and goes to the state 
  $(p',S_P\cup S_{p,a,p'})$, where $S_P$ are the successor states of
  of $P$ by $\delta$ on the input $a$, and $S_{p,a,p'}$ are all the
  states reached from $p$ by a transition on $a$ bigger than
  $(p,q,p')$ (according to $<_\delta$).

  Formally, $((p,P),a,(p',P'))\in\delta'$ iff 

  \begin{itemize}
    \item $(p,a,p')\in\delta$ ;
    \item $P' = \{ q'\ |\ \exists q\in P,\ (q,a,q')\in\delta\}\cup \{
      p''\ |\ (p,a,p'')\in\delta\wedge (p,a,p')<_{\delta} (p,a,p'')\}$.

  \end{itemize}

  It is clear by construction that $A$ and $A'$ define the same
  domain. As $A$ is functional, they also define the same function,
  because the value of a word is equal to the value of any run on it,
  and in particular to the value of the smallest run (w.r.t $<_\delta$).
\end{proof}

\newcommand{\neutral}{\mathbb{1}}

\vspace{2mm}
\noindent \textbf{Group Automata} Whenever possible, we generalise our result
to group automata. Group automata are defined as weighted automata,
except that the weights are taken from a group structure. Since there
are no operation to combine the values of all the runs on the same
input word, group automata define functions from words to set of
values. Let us recall the definition of a group. A \emph{group} 
is a structure $(W,\cdot,\neutral)$,  where $W$ is a set, $\cdot:W\times
W\rightarrow W$ is an associative operation, $\neutral\in W$ is a two sided
identity element for $\cdot$ over $W$, and each element $s\in W$
admits an inverse $ s^{-1}\in W$, such that $s^{-1}\cdot s= s\cdot
s^{-1}=\neutral$ (the inverse is unique).

A \emph{group automaton} over $\Sigma$ and a group
$(W,\cdot,\neutral)$ is a tuple $A {=} (Q,q_I,F,\delta,\gamma)$ where 
$Q,q_I,F,\delta$ are defined as for weighted automata, and 
$\gamma : \delta\rightarrow W$ is the weight function. 
Runs are defined as for weighted automata, and the value $V(\rho)$ of a
run is obtained by taking the product of the weights of the
transitions of $\rho$. The language defined by $A$, denoted by $L_A$,
is a function from $\Sigma^+$ to finite subsets of $W$, defined by 
$L_A(w) = \{ V(\rho)\ |\ \rho\text{ is an accepting run on } w\}$. 

All the notions defined for weighted automata, such as functionality,
accessible states, etc. carry over to group automata. When a group
automaton $A$ is functional, we rather write $L_A(w) = v$ instead of 
$L_A(w) = \{ v\}$. Finally, we
assume that the operations over group elements are computable
in polynomial time in the size of their representation.

When considering only the relation induced by weighted $V$-automata,
it turns out that they are equivalent to group automata, for
$V\in\{\Sum,\Avg,\Dsum\}$. 

\begin{lem}\label{lem:encodegroup}
Let $V\in\{\Sum,\Avg,\Dsum\}$. There exists a group $G_V$ and a partial
function $\psi : G_V\rightarrow \mathbb{Q}$, such that 
for all weighted $V$-automaton $A$, one can construct, in linear-time,
a group automaton $A_g$ over $G_V$ such that for all $w\in\Sigma^+$, we
have:
\begin{itemize}
    \item for all $v\in\mathbb{Q}$, if $(w,v)\in R_A^V$,
      there exists $s\in G_V$ s.t. $s\in L_{A_g}(w)$ and $\psi(s) = v$,
    \item for all $s\in G_V$ s.t. $s\in L_{A_g}(w)$,
      $(w,\psi(s))\in R_A^V$.
\end{itemize}
\end{lem}

\begin{proof}
For \Sum and \Avg, we  define $G_\Sum = ( \mathbb{Z},+, 0)$ and 
$G_\Avg = (\mathbb{Z}^2,\cdot,(0,0))$, where $\cdot$ is the pairwise
sum. An \Avg-automaton $A$ can be turned into a group automaton over
$G_\Avg$ by replacing each weight $v$ in $A$ with the pair
$(v,1)$. Then, it suffices to take $\psi$ defined by $\psi((v,n)) =
v/n$ for all $n\geq 1$, otherwise $\psi$ is undefined.

For \Dsum-automata, consider the group  $G_\Dsum = (W,\cdot,\neutral)$, where
$W=\mathbb{Q}\times \mathbb{Q}^+$ (where $\mathbb{Q}^+$ is the set of
strictly positive rational numbers), $\cdot$ is defined by
$(a,x)\cdot(b,y)=(\frac{1}{y}a+b,xy)$, $\neutral = (0,1)$ is the identity
element, and given $(a,x)\in W$, the inverse $(a,x)^{-1}$ is given by
$(a,x)^{-1}=(-xa,x^{-1})$.

Given $\lambda\in\mathbb{Q}\cap ]0,1[$,  a  \Dsum-automaton $A$ on
$\Sigma$  can be turned into a group automaton on $G_\Dsum$,  by
replacing each weight $a$ in $A$ with the  pair $(a,\lambda),
a\in\mathbb{Z}$. Let $w=w_0\dots w_n\in\Sigma$, and consider a run
$\rho:q_0\run{w} q_{n+1}$ on $w$ in $A$. Then,  $\rho$  is valued by
the pair $(a,x)=(\frac{1}{\lambda^{n}}\gamma(q_0,w_0,q_1)+\dots +
\gamma(q_{n},w_n,q_{n+1}),\lambda^{n+1})$. Hence,  $(a,x)$ codes the
value $\psi((a,x)) = \frac{ax}{\lambda}=\Dsum(\rho)$.
\end{proof}

As an immediate consequence of Lemma \ref{lem:encodegroup}, we obtain that $A$ is functional iff
$A_g$ is, and therefore, any procedure to decide functionality of
group automata will allow one to decide functionality of weighted
$V$-automata, for $V\in \{\Sum,\Avg,\Dsum\}$. We show such a procedure
in the next section. We leave open the encoding of \Ratio-automata as
group automata.

\section{Functionality Problem}
In this section, we consider the problem of deciding whether a weighted automaton  is functional. 
In particular, Subsection \ref{sec:func1} provides a general
functionality test applying to group automata which, by Lemma
\ref{lem:encodegroup}, yields a procedure to test functionality of
weighted $V$-automata, for $V\in \{\Sum,\Avg,\Dsum\}$. %
We also prove that functionality is decidable for \Ratio-automata, by
using different techniques.

\subsection{Functionality of group automata}\label{sec:func1}
We start to introduce the notion of \emph{delay} between two runs of a
group automaton, which turns out to be the main ingredient of the
functionality algorithm in Fig.~\ref{alg:funcTest}.

%
%
%

\begin{defi}[Delay] \label{defDelay} Let $G = (W,\cdot,\neutral)$ be a
    group, and $A = (Q,q_I,F,\delta,\gamma)$ be a group automaton over
    $G$. Let $p,q\in Q$. A value $d\in W$ is a \emph{delay} for $(p,q)$ if
    $A$ admits two runs $\rho:q_I\run{w}p$,
    $\rho':q_I\run{w}q$ on $w\in\Sigma^*$ s.t. $\delay(\rho,\rho')
    {=_{def}} (V(\rho))^{-1}\cdot V(\rho')=d$.
\end{defi}

\noindent The following lemma shows that  at most one delay can be associated with
co-accessible pairs of states in a functional group automaton. This is related
to the uniqueness of inverse elements. 

\begin{lem}[One Delay] \label{onedelay} Let $A =
(Q,q_I,F,\delta,\gamma)$ be a functional group automaton.
For all pairs of states $(p,q)$: If $(p,q)$
is co-accessible, then $(p,q)$ admits at
most one delay.
\end{lem}
\begin{proof}
Let $(p,q)$ be a co-accessible pair of states. Assume that $(p,q)$ admits two delays $d_1,d_2$. We show that if $A$ is functional, then $d_1=d_2$. Let $\rho_1:q_0\run{w_1|n_1} p$, $\rho'_1:q_0\run{w_1|m_1} q$ (resp. $\rho_2:q_0\run{w_2|n_2} p$, $\rho'_2:q_0\run{w_2|m_2} q$) be a run witnessing the delay $d_1=\delay(\rho_1,\rho'_1)$ (resp. $d_2=\delay(\rho_2,\rho'_2)$).  Since $(p,q)$ is co-accessible, there exists a word $u$ and two runs 
$\rho_3:p\run{u|l} f\in F$, $\rho'_3:q\run{u|s} f'\in F$. The hypothesis of functionality yields:
\begin{gather}
m_1\cdot s = n_1\cdot l\Rightarrow n_1^{-1}\cdot m_1= l\cdot s^{-1}\label{eqn:1}\\
n_2\cdot l = m_2\cdot s \Rightarrow l\cdot s^{-1}= n_2^{-1} \cdot m_2\label{eqn:2}
\end{gather}
From the above equations, we get:
\begin{equation}
 n_1^{-1}\cdot m_1= l\cdot s^{-1}= n_2^{-1}\cdot m_2 \Rightarrow  m_1^{-1}\cdot n_1=  m_2^{-1}\cdot n_2\label{eqn:3}
\end{equation}

\end{proof}

\noindent We are now ready to define an algorithm (Algorithm \ref{alg:funcTest})
 that checks the functionality of a group automaton over a group $(W,\cdot,\neutral)$. In
 a first step, such a procedure computes all co-accessible pairs of
states. Then, it explores the set of accessible pairs of states in
a forward manner and computes the delays associated with those pairs.
If two different delays are associated with the same pair, or if a pair of
final states with a  delay different from $\neutral$ (the neutral element of the group) is reached, the test stops
and returns that the automaton is not functional (by Lemma \ref{onedelay} and by definition of functionality). Otherwise, it goes on
until all co-accessible (and accessible) pairs have been visited and
concludes that the automaton is functional. 

\begin{algorithm}[h]
\caption{Functionality test for group automata.\label{alg:funcTest} }

{
\textbf{Input}: A group automaton $A =
 (Q,q_I,F,\delta,\gamma)$ over a group $G = (W,\cdot,\neutral)$.
\\
\textbf{Output}: {\sf Yes} if $A$ is functional, {\sf No} otherwise.
\\
\Begin{
\nl ${\sf CoAcc} \gets$ all co-accessible pairs of states\;
\nl ${\sf visited} \gets \emptyset$ \label{alg1:line1}; ${\sf
delay}(q_I,q_I)\gets \neutral; {\sf PUSH}(S,((q_I,q_I),(\neutral,\neutral)))$ \;

\nl \While{$S \neq \emptyset$ \label{alg1:while-loop}}
{
\nl $((p,q),(\alpha,\beta)) \gets {\sf POP}(S)$\; \label{alg1:pop}
\nl \lIf{$(p,q)\in F^{2} \wedge \alpha^{-1}\cdot\beta\neq \neutral$\label{alg1:testFinal}} {${\bf returns}\ \  {\sf No}$}\;
\nl \eIf{$(p,q)\in {\sf visited}$}{
 \lIf{$\delay(p,q)\neq \alpha^{-1}\cdot\beta$ \label{alg1:test2delays}}{$ {\bf returns}\ \  {\sf No}$}}
{
  \nl ${\sf visited } \gets {\sf visited}\cup \{(p,q)\}$\;
  \nl $ {\sf delay}(p,q)\gets \alpha^{-1}\cdot\beta$\;
  \nl  \lForEach{$(p',q')\in{\sf CoAcc}$ s.t. $\exists a\in\Sigma\cdot 
 (p,a,p')\in\delta\wedge(q,a,q')\in\delta$}{${\sf PUSH}(S,((p',q'),(\alpha\cdot\gamma(p,a,p'),\beta\cdot\gamma(q,a,q')))$ \;}
}
}
\nl ${\bf returns}\ \ {\sf Yes}$
}
}
\end{algorithm}


If the algorithm returns  {\sc No}, it is either because
a pair of accepting states with a delay different from $1$  has been reached, which gives
a counter-example to functionality, or because  a pair of states with
 different delays has been found, so $A$ is not functional by
Lemma \ref{onedelay}.

Lemma \ref{lemmapumping} provides a pumping argument useful to establish the converse. In detail, given a non functional automaton $A$,  Lemma \ref{lemmapumping} establishes  that  $A$  admits
two accepting runs witnessing non-functionality (i.e. on the same word $w$ and with different values)  for which  any pair of states that repeats twice  has two different delays.  The proof of   Lemma \ref{lemmapumping} relies on  the following key result:
\begin{lem}\label{lem:removeslice}
Let $A=(Q,q_I,F,\delta,\gamma)$ be a group automaton. Let
$w_1,w_2,w_3\in\Sigma^*$ such that there exist $p,p',q,q'\in Q$ and
the following runs:
$$
\begin{array}{lllllll}
  \rho_1: q_I\xrightarrow{w_1|n_1} p & \rho_2: p\xrightarrow{w_2|n_2}p &
  \rho_3 : p\xrightarrow{w_3|n_3} q \\
  \rho_1': q_I\xrightarrow{w_1|m_1} p' & \rho_2: p'\xrightarrow{w_2|m_2}p' &
  \rho_3 : p'\xrightarrow{w_3|m_3} q' \\
\end{array}
$$
s.~t.~$\delay(\rho_1,\rho'_1) {=}
\delay(\rho_1\rho_2,\rho'_1\rho'_2)$. 
Then, we have $\delay(\rho_1\rho_2\rho_3,\rho'_1\rho'_2\rho'_3) {=}
\delay(\rho_1\rho_3,\rho'_1\rho'_3)$.
\end{lem}
\begin{proof}
By hypothesis, we have:
\begin{gather}
n_1^{-1}\cdot m_1 = (n_1\cdot n_2)^{-1}\cdot m_1\cdot m_2 \Rightarrow 
n_1^{-1}\cdot m_1= n_2^{-1}\cdot n_1^{-1}\cdot m_1\cdot m_2 \notag
\end{gather}
Therefore, $n_3^{-1}\cdot n_1^{-1}\cdot m_1\cdot m_3 =  n_3^{-1}\cdot n_2^{-1}\cdot n_1^{-1}\cdot m_1\cdot m_2\cdot m_3$, that implies $(n_1\cdot n_3)^{-1}\cdot (m_1 \cdot m_3)= (n_1\cdot n_2\cdot n_3)^{-1}\cdot m_1\cdot m_2 \cdot m_3$ i.e. $\delay(\rho_1\rho_2\rho_3,\rho'_1\rho'_2\rho'_3) =
\delay(\rho_1\rho_3,\rho'_1\rho'_3)$. 
\end{proof}
\begin{lem}\label{lemmapumping}  
Let $A {=} (Q,q_I,F,\delta,\gamma)$ be a group automaton.
 If $A$ is not functional, there exists a
word $w{=}\sigma_0\dots\sigma_n$ and two accepting runs on it,
$\rho{=}q_0\sigma_0\dots q_{n+1}$, $\rho'{=}q'_0\sigma_0\dots q'_{n+1}$, such
that $V(\rho){\neq}V(\rho')$ and for
all positions $i < j$ in $w$, if $(q_i,q'_i) {=}  (q_j,q'_j)$ then 
$\delay(\rho_i,\rho'_i){\neq} \delay(\rho_j,\rho'_j)$. 
\end{lem}
\begin{proof}
Let $w{=}\sigma_0\dots\sigma_n$, $w\in\dom(A)$ such that $|R_A(w)|>1$. Clearly, there exist
two runs $\rho{=}q_0\sigma_0\dots q_{n+1},\rho'{=}q'_0\sigma_0\dots q'_{n+1}$  on $w$ such that
$V(\rho)\neq V(\rho')$. This implies $\delay(\rho,\rho')\neq 1$. Suppose that there are two positions $i<j$  such that $(q_i,q'_i)=(q_j,q'_j)$ and $\delay(\rho_i,\rho'_i)=\delay(\rho_j,\rho'_j)$. Then, Lemma \ref{lem:removeslice}  applies and we can shorten the runs $\rho,\rho'$ by removing the slices starting from position $i$ and ending in position $j$. In particular, we obtain two new runs $\bar{\rho}{=}q_0\sigma_0\dots q_i \sigma_{j+1}\dots  q_{n+1}$, $\bar{\rho}'{=}q'_0\sigma_0\dots q_i\sigma_{j+1}\dots  q'_{n+1}$  on $\bar{w}=\sigma_0\dots\sigma_{i-1}\sigma_{j+1}\dots\sigma_n$ with the following property: $\delay(\bar{\rho},\bar{\rho}')=\delay(\rho,\rho')\neq 1$. Iterating the above procedure we recover a pair of runs $(\widehat{\rho},\widehat{\rho}')$ on $\widehat{w}$ such
that $\delay(\widehat{\rho},\widehat{\rho}')\neq \neutral$ and for
all positions $i < j$ in $\widehat{w}$, if $(q_i,q'_i) {=}  (q_j,q'_j)$ then 
$\delay(\widehat{\rho}_i,\widehat{\rho}'_i){\neq} \delay(\widehat{\rho}_j,\widehat{\rho}'_j)$, fulfilling the goals of Lemma   \ref{lemmapumping}.
\end{proof}

If there are two runs witnessing non-functionality without repetitions of 
pairs of states, the algorithm can find a pair of final
states with a  delay different from $\neutral$. Otherwise the algorithm will return {\sc No} at line \ref{alg1:test2delays}, if not before. Therefore we get:



\begin{thm} \label{teocorsum}  
Let $A = (Q,q_I,F,\delta,\gamma)$ be a group automaton. Algorithm \ref{alg:funcTest}  returns  {\sc Yes} on $A$ iff $A$  is functional and terminates within
$O(|A|^2)$ steps.
\end{thm}
\begin{proof} 
The procedure explores the co-accessible part of $A\times A$ in a depth-first-search manner. In particular, unless a condition for early termination (answering {\sc No}) is encountered, each edge of the (co-accessible and accessible part of) $A\times A$ is visited once.

Given the above premise, assume that $A$ is not functional. Then by Lemma \ref{lemmapumping}, $A$ admits two runs witnessing non functionality for which any pair of states that repeat twice have different delays. Thus, the exploration of such a pair of parallel runs in Algorithm \ref{alg:funcTest}  will eventually stop the procedure returning {\sc No}, as soon as either a pair of co-accessible states is visited twice with different delays, or the final states are reached with a delay different from $\neutral$. Conversely, assume that the algorithm returns {\sc No}. Then, this happens either at line $5$ or at line $6$. In the first case, the algorithm exhibits exactly a pair of runs witnessing non-functionality (by definition). In the second case, the answer is correct by Lemma \ref{onedelay}. 


The complexity follows from the following observations. At line $1$ the set of co-accessible pairs of states can be obtained in 
$O(|A|^2)$ steps by proceeding as follows. First, compute $A\times A$ and reverse its edges. Then for each pair $(q,q')$, $q\in F,q'\in F$, mark it as co-accessible and use it to discover new co-accessible pairs of states via a depth-first-search visit on (non-marked) states in $A\times A$. Finally, within the main loop at line $3$ each co-accessible pair of states is inserted  into $S$ (with a corresponding delay) at most once. 
\end{proof}
 
\begin{rem}
Functionality of \Sum-automata have been shown decidable in
\cite{KirstenM05}. Our functionality algorithm on weighted automata
over groups specialized for \Sum-automata  corresponds to the
functionality algorithm  for \Sum-automata defined in
\cite{KirstenM05}.  Algorithm \ref{alg:funcTest} can also be applied
to word transducers for which functionality has been shown to be decidable in PTIME with similar techniques in \cite{BealEtAl03a}.
\end{rem}

Corollary \ref{corfunc} follows immediately from Lemma \ref{lem:encodegroup} and Theorem \ref{teocorsum}. 

\begin{cor}\label{corfunc}
      The functionality problem is decidable in \ptime for $V$-automata, $V\in\{\Sum,\Avg,\Dsum\}$.        
\end{cor}
%
\begin{rem}\label{remarkCong} We remark that our functionality test could be applied also to a more general framework, where the weight-set $W$ of the considered group automaton is equipped with an equivalence relation $\sim_W$, and two runs are considered equivalent iff the corresponding values are equivalent w.r.t. $\sim_W$. In particular, $\sim_W$ needs to fulfill the following properties in order to be able to show uniqueness of the delay (modulo $\sim_W$) and termination of the functionality test: $(1)$ it is a \emph{congruence}, i.e. $\forall a,b,c,d\in W$ if $a\sim_Wb$ and $c\sim_W d$, then $a\cdot c\sim_W b\cdot d$; $(2)$ for all $a,b,c\in W$ if $a\nsim_W b$ then $a\cdot c\nsim_W b\cdot c$.
\end{rem}

\subsection{Functionality of \Ratio-automata}\label{sec:Ratio}

        Unlike \Sum, \Avg and \Dsum-automata, it is unclear whether \Ratio-automata can
    be encoded in term of group automata. Intuitively, to provide such an encoding we would assign to each edge a pair of natural numbers, where the first component is the edge-reward and the second component is the edge-cost. Thus, each run $\rho$ is assigned the value $(n,m)$, where $n$ (resp. $m$) is the sum of the rewards (resp. costs) along the run, and two runs $\rho,\rho'$ with values $(n,m), (n',m')$ need to be considered equivalent iff $nm'=n'm$. Unfortunately,  the induced  equivalence relation (where $(n,m)$ is equivalent to $(n',m')$ iff $nm'=n'm$) is not a congruence.
        Therefore, the results developed in the previous subsection do not apply to this class of weighted automata (at least to this encoding) as the quotient of the set of pairs by this equivalence relation is not a group  (cf. Remark \ref{remarkCong}). In fact,
         it is still open whether
        there exists a good notion of delay for \Ratio-automata
        that would allow us to design an efficient algorithm to test
        functionality. However deciding functionality  can be done
        by using a short witness property of non-functionality, as shown in the following lemma.

\begin{lem}[Pumping]\label{lemmaPumpingRatio}
  Let $A$ be a \Ratio-automaton with $n$ states. $A$ is not functional
  iff there exist a word $w$ such that $|w|< 4n^2$ and two accepting runs $\rho,\rho'$ on $w$ such that
  $\Ratio(\rho)\neq \Ratio(\rho')$.
\end{lem}

\begin{proof}
  We prove the existence of a short witness for non-functionality. The
  other direction is obvious. Let $w$ be a word such that $|w|\geq 4n^2$ 
  and there exist two accepting runs $\rho_1,\rho_2$ on $w$ such that
  $\Ratio(\rho)\neq \Ratio(\rho')$. Since $|w|\geq 4n^2$, there exist states
  $p,q\in Q, p_f,q_f\in F$ and words $w_0,w_1,w_2,w_3,w_4$ such that $w=w_0w_1w_2w_3w_4$
  and $\rho,\rho'$ can be decomposed as follows:
  $$
  \begin{array}{llllllllllllllllllllllllll}
    \rho: &q_I&\xrightarrow{w_0|(r_0,c_0)}&
    p&\xrightarrow{w_1|(r_1,c_1)}&p&\xrightarrow{w_2|(r_2,c_2)}&p&\xrightarrow{w_3|(r_3,c_3)}&p&\xrightarrow{w_4|(r_4,c_4)}&p_f
    \\
    \rho': &q_I&\xrightarrow{w_0|(r'_0,c'_0)}&
    q&\xrightarrow{w_1|(r'_1,c'_1)}&q&\xrightarrow{w_2|(r'_2,c'_2)}&q&\xrightarrow{w_3|(r'_3,c'_3)}&q&\xrightarrow{w_4|(r'_4,c'_4)}&q_f
    \\
  \end{array}
  $$
  where $r_i,c_i$ denotes the sum of the rewards and the costs
  respectively on the subruns of $\rho$ on $w_i$, and similarly for
  $r'_i,c'_i$.

  By hypothesis we know that $(\sum_{i=0}^4 r_i)\cdot (\sum_{i=0}^4
  c'_i)\neq (\sum_{i=0}^4 c_i)\cdot (\sum_{i=0}^4 r'_i)$. For all
  subsets $X \subseteq \{1,2,3\}$, we denote by $w_X$ the word
  $w_0w_{i_1}\dots w_{i_k}w_4$ if $X = \{ i_1 < \dots < i_k\}$. For
  instance, $w_{\{1,2,3\}} = w$, $w_{\{1\}} = w_0w_1w_4$ and $w_{\{\}}
  = w_0w_4$. Similarly, we denote by $\rho_X,\rho'_X$ the
  corresponding runs on $w_X$. We will show that there exists 
  $X\subsetneq \{1,2,3\}$ such that $\Ratio(\rho_X)\neq \Ratio(\rho'_X)$.
  Suppose that for all $X\subsetneq \{1,2,3\}$, we have
  $\Ratio(\rho_X) =  \Ratio(\rho'_X)$. We now show that it implies that
  $\Ratio(\rho) = \Ratio(\rho')$, which contradicts the hypothesis.
  For all $X\subseteq \{1,2,3\}$, we let:
  $$
  L_X\ =\ (\sum_{i\in X\cup\{0,4\}} r_i)\cdot (\sum_{i\in
    X\cup\{0,4\}} c'_i)\qquad R_X\ =\ (\sum_{i\in X\cup\{0,4\}} c_i)\cdot (\sum_{i\in
    X\cup\{0,4\}} r'_i)
  $$
  By hypothesis, $L_{\{1,2,3\}}\neq R_{\{1,2,3\}}$ and for all
  $X\subsetneq \{1,2,3\}$, $L_X = R_X$. The following equalities can be easily verified:
  \vspace{-2mm}
  $$
  \begin{array}{llllllllllllllllllllllll}
    L_{\{\}} &+& L_{\{1,2\}} &+& L_{\{1,3\}} &+& L_{\{2,3\}} &-& L_{\{1\}} &-&
    L_{\{2\}} &-& L_{\{3\}} & = & L_{\{1,2,3\}} \\
    R_{\{\}} &+& R_{\{1,2\}} &+& R_{\{1,3\}} &+& R_{\{2,3\}} &-& R_{\{1\}} &-&
    R_{\{2\}} &-& R_{\{3\}} & = & R_{\{1,2,3\}} \\
  \end{array}
  $$


    \noindent Then, since by hypothesis we have $L_X=R_X$ for all
    $X\subsetneq \{1,2,3\}$, we get $L_{\{1,2,3\}} = R_{\{1,2,3\}}$,
    which is a contradiction. Thus there exists $X\subsetneq
    \{1,2,3\}$ such that $L_X\neq R_X$. In other words, there exists
    $X\subsetneq \{1,2,3\}$ such that $\Ratio(\rho_X)\neq \Ratio(\rho'_X)$. 
    This shows that when a witness of non-functionality has length at
    least $4n^2$, we can find a strictly smaller witness of
    functionality.  
\end{proof}

\begin{thm}\label{thm:funratio}
  Functionality is decidable in {\sf CoNP} for \Ratio-automata, and in
  {\sf PTime} if the rewards and costs are unary encoded.
\end{thm}
\begin{proof} 
As a consequence of Lemma \ref{lemmaPumpingRatio}, we can design an {\sf NP}
procedure that will check non-functionality by guessing runs of length
at most $4n^2$ and by testing, in polynomial time, that they have
different ratio values, where $n$ is the number of states.

To get the {\sf PTime} upper bound, we design a {\sf CoNLogSpace} algorithm
as follows. The algorithm guesses two runs $\rho_1$ and $\rho_2$ of the
\Ratio-automaton $A$ on the same input, and stores in memory a counter value $c$, the
two states $q_1$ and $q_2$ reached so far by the runs, and the current
sums $R_1,C_1,R_2,C_2$ of costs and rewards of the two runs
respectively. At each step, it guesses a new symbol $\sigma\in\Sigma$,
two transitions $q_1\xrightarrow{\sigma|(r_1,c_1)} q'_1$ and
$q_2\xrightarrow{\sigma|(r_2,c_2)} q'_2$ of $A$, and updates its
memory to $(c+1,q'_1,q'_2,R_1+r_1,C_1+c_1,R_2+r_2,C_2+c_2)$. The
algorithm stops in a configuration $(c,q_1,q_2,R_1,C_1,R_2,C_2)$ with
negative answer whenever $c = 4n^2+1$, 
and with a positive answer whenever $c\leq 4n^2$, $q_1,q_2$ are accepting,
and $R_1C_2\neq R_2C_1$.

The correctness of this algorithm directly
follows from Lemma \ref{lemmaPumpingRatio}. Let us show that each
configuration uses only a logarithmic space to be stored. Let $M$ be
the largest value (among rewards and costs) that occurs on the
transitions of $A$. Let $|M|$ denote the size of $M$. Since rewards
and costs are encoded in unary, we have $|M| = M$. Now, the algorithm
first converts the rewards and costs in binary (in logspace), and
then, each configuration $(c,q_1,q_2,R_1,C_1,R_2,C_2)$ takes only logarithmic
space, because for any $\alpha\in\{R_1,C_1,R_2,C_2\}$, $\alpha\leq
4n^2M = 4n^2|M|$, and therefore $log(\alpha)\leq 2log(2n)+log(|M|)$. 
We can conclude since {\sf CoNLogSpace}$\subseteq${\sf PTime}.
\end{proof} 
\begin{rem}
        The pumping result in Lemma \ref{lemmaPumpingRatio} states that if some \Ratio-automaton with
        $n$ states is not functional, there exists a witness of non-functionality whose length
is bounded by $4n^2$, where $n$ is the number of states. 
Such a property also holds for \Dsum-automata (and is well-known
for \Sum and \Avg-automata), but with the smaller bound $3n^2$.
Those bounds are used to state the existence of two runs on the same
word such that the same pair of states is repeated 3 or 4 times along
the two runs. Then it is proved that one can remove some part in between
two repetitions and get a smaller word with two different output values.
However for \Ratio-automata, three repetitions are not enough to be
able to shorten non-functionality witnesses. For instance, consider
the following two runs on the alphabet $\{a,b,c,d\}$ and states 
$\{q_I,p,q,p_f,q_f\}$ where $p_f,q_f$ are final (those two runs can
easily be realized by some \Ratio-automaton):
  $$
  \begin{array}{llllllllllllllllllllllllll}
    \rho: &q_I&\xrightarrow{a|(2,2)}&
    p&\xrightarrow{b|(2,1)}&p&\xrightarrow{c|(2,2)}&p&\xrightarrow{d|(1,1)}&p_f
    \\
    \rho': &q_I&\xrightarrow{a|(1,2)}&
    q&\xrightarrow{b|(2,1)}&q&\xrightarrow{c|(1,1)}&q&\xrightarrow{d|(2,1)}&q_f
    \\
  \end{array}
  $$
It is easy to verify that the word $abcd$ has two outputs given by
$\rho$ and $\rho'$ while the
words $ad$, $abd$ and $acd$ has one output. For instance, the two
runs $q_I\xrightarrow{a|(2,2)} p\xrightarrow{d|(1,1)}p_f$ and 
$q_I\xrightarrow{a|(1,2)}q\xrightarrow{d|(2,1)}q_f$ on $ad$ have both
value $1$.
\end{rem}

\section{Verification Problems \label{secdec}}
In this section, we investigate several decision problems for
functional $V$-automata as defined in \cite{ChatterjeeDH10}, $V\in\{\Sum,\Avg,\Dsum,\Ratio\}$. Given two $V$-automata $A,B$ over
$\Sigma$ (and with the same discount factor when $V = \Dsum$) and a threshold $\nu\in\mathbb{Q}$, we define the following decision problems, where $\triangleright\in\{>,\geq\}$:

%
%
%
%
%
%
\vspace{2mm}
\begin{tabular}{@{\hspace{-4mm}}llllll}
{\sf $\triangleright \nu$-Emptiness} & $L_A^{\triangleright \nu}\neq\varnothing$ & holds if there exists
$w\in\Sigma^+$ such that $L_A(w)\triangleright \nu$\\

{\sf $\triangleright \nu$-Universality} & $L_A\triangleright \nu$ & holds if for all
$w\in\dom(A)$,  $L_A(w)\triangleright \nu$. \\

{\sf Inclusion} & $L_A\leq L_B$ & holds if for all $w\in\Sigma^+$,
$L_A(w)\leq L_B(w)$\\

{\sf Equivalence} & $L_A =  L_B$ & holds if for all $w\in\Sigma^+$,
$L_A(w) = L_B(w)$\\
\end{tabular}
\vspace{2mm}

\noindent Theorem \ref{teoEmpty} (resp. Theorem\ref{teoUniversal}) below proves that the $\triangleright \nu$-emptiness (resp.  universality) problem can be solved in polynomial time for functional 
\Sum-, \Avg-, and \Ratio-automata. As for functional \Dsum-automata, we provide a \ptime upper bound for the ${>}\nu$-emptiness (resp. ${\geq}\nu$-universality)  problem. The ${\geq} \nu$-emptiness  (resp. ${>} \nu$-universality)  on functional  \Dsum-automata has been recently shown in  {\sf PSpace}  in \cite{boker1}, as an aside result of a partial solution\footnote{In particular, \cite{boker1} provides a solution to the finite version of the target discounted-sum problem.} to the notorious target discounted-sum problem. The latter is known to be connected to various areas and open problems in mathematics and computer science, such as e.g. the universality of (nondeterministic) \Dsum-automata, \Dsum-games with imperfect information or multi-objectives \cite{DBLP:conf/csl/BokerH11,Chatterjee13}, as well as  piecewise affine maps and generalizations of the Cantor set \cite{boker1}.
\begin{thm}\label{teoEmpty}
   The $\triangleright \nu$-emptiness problem is in \ptime
 for functional \Sum-, \Avg-, and \Ratio-automata. The ${>}\nu$-emptiness  (resp.
 ${\geq} \nu$-emptiness) problem is in \ptime (resp. {\sf PSpace} \cite{boker1}) for functional \Dsum-automata.
\end{thm}

\begin{proof}
For $\Sum$-automata, let $A = (Q,q_I,F,\delta,\gamma)$ be a \Sum-automaton. Wlog we assume that all states of $A$ are both accessible from an initial state and co-accessible from a final state (such property can be ensured via a \ptime transformation \cite{Mohri09}). First, $L_A^{\triangleright \nu}\neq \varnothing$ if $A$ contains a strictly positive cycle, otherwise one inverts the weights and computes a shortest path from an initial to a final state. If the sum $\beta$ of such a path satisfies ${-}\nu \triangleright \beta$ then the language is non-empty. Both steps are handled by the classical Bellman-Ford algorithm. 

For \Avg-automata, let $A = (Q,q_I,F,\delta,\gamma)$ be an \Avg-automaton. We can assume $\nu=0$ since the $\triangleright \nu$-emptiness problem for \Avg-automata reduces to the $\triangleright 0$-emptiness problem for \Sum-automata, by simply reweighting the input automaton. Then, $L_A^{\triangleright 0}\neq\emptyset$ iff $A$ admits a path to a final state whose sum of the weights is $\triangleright 0$, that  can be easily checked in \ptime.

For \Dsum automata,  let $A = (Q,q_I,F,\delta,\gamma)$ be a \Dsum-automaton. We show that $L_A^{>\nu}\neq\emptyset$ iff Player $0$ has a strategy to ensure a play from $v_I$ with discounted sum greater than $\nu$   in the one player (infinite) \Dsum game $\Gamma=(V, E, w, \langle V_0,V_1\rangle)$, where:
\begin{itemize}
\item $V=\{p\:|\:p\in Q\wedge \exists w\in \Sigma^* (p\overset{w}{\rightsquigarrow}f\in F)\}$
\item $V_0=V, V_1=\emptyset$
\item $E=(V\times (\Sigma\cup\{\zeta\})\times V)\cap(\{(p,a,p')\:|\:(p,a,p')\in \delta\}\cup\{(p,\zeta,p)\:|\:p\in F\})$, where $\zeta\notin\Sigma$ is a fresh symbol
\item For each  $e=(p,a,p')\in E$: If $(p,a,p')\in \delta$, then $w(e)=\gamma(p,a,p')$, else $w(e)=0$.
\end{itemize}
Once proved the above equivalence, our complexity bound follows easily, since checking whether $L_A^{>\nu}\neq\emptyset$ reduces to  solving  a $1$ player \Dsum game (that is in \ptime \cite{andersson2006}).

 If $L_A^{>\nu}\neq\emptyset$, then  $A$  admits an accepting run $\rho$ such that  $\Dsum(\gamma(\rho))>\nu$. By construction, $\Gamma$ admits an (infinite) path  with a positive discounted sum, i.e. Player $0$ has a (memoryless) strategy to win the one-player discounted sum game $\Gamma$.  

Conversely, suppose that Player $0$ has a strategy to win the one-player discounted sum game $\Gamma$. Let $r$ be an infinite path on $\Gamma$ consistent with a winning strategy for player $0$. Then $\Dsum(r)>0$. Let $W$ be the maximum absolute weight in $\Gamma$. For each prefix $r_i$ of length $i$
of $r$ we have: 
\begin{equation}\label{eqcomplan1} 
\begin{split}
&\Dsum(r_i)+\dfrac{W\lambda^i}{1-\lambda}\geq\Dsum(r)\Rightarrow\\
&\Dsum(r_i)\geq\Dsum(r)-\dfrac{W\lambda^i}{1-\lambda}
\end{split}
\end{equation}
Since $\Dsum(r)>\nu$, there exists $i^*$ such that
$\Dsum(r)-\dfrac{W\lambda^{i^*}}{1-\lambda}>\nu$ that implies
$\Dsum(r_{i^*})>\nu$.  By construction, each path in $\Gamma$ can be
extended to reach a node in $F$. Let $r'_{i}=r'_0\dots r'_m\in F$ be
such a continuation of $r^{i^*}$. By Equation  \ref{eqcomplan1}, our choice of $i^*$ guarantees that $\Dsum(r'_i)>\nu$. Since $A$ is functional, $r'$ witnesses the existence of a word $w$ such that $L_A(w)>\nu$.

Finally, let $A$ be a \Ratio-automaton, let $\nu=m/n$. We consider the \Sum automaton $A'$, where each edge of $A$ having reward $r$ and cost $c$ is replaced by an edge of weight $rn-cm$. It can be easily proved that   $L_A^{\triangleright \nu}\neq\emptyset$ iff $L_{A'}^{\triangleright \nu}\neq\emptyset$.
\end{proof}


\begin{thm}\label{teoUniversal}
Let $\nu\in\mathbb{Q}$. The $\triangleright \nu$-universality problem is {\sf PTime}  for functional  $V$-automata, $V\in \{\mbox{\Sum, \Avg, \Ratio}\}$.
The ${\geq} \nu$-universality  (resp.  ${>} \nu$-universality) problem is {\sf PTime}  (resp. {\sf PSpace} \cite{boker1}) for functional  \Dsum-automata.
\end{thm}
\begin{proof}
 Let $A$ be a $V$-automaton, $V\in\{\Sum,\Avg,\Dsum\}$ and consider
 the $\geq \nu$-universality  (resp.  $> \nu$-universality) problem
 for $V$-automata. We check whether $A$ admits an
  accepting run with $V(\gamma(r)) <\nu$ (resp. $V(\gamma(r)) \leq \nu$) . This can be done in \ptime
  for $V\in\{\Sum, \Avg, \Ratio,  \Dsum\}$
  (resp. $V\in\{\Sum, \Avg, \Ratio\}$), with a procedure similar to
  the one applied in the proof of Theorem \ref{teoEmpty}.
\end{proof}

It is known that inclusion is undecidable
for non-deterministic \Sum-automata \cite{Krob/94,AlmagorBokerKupferman2011}, and therefore also for \Avg and \Ratio-automata. To the best of our
knowledge, it is open whether it is decidable for \Dsum-automata. This situation is strikingly different for functional automata as the inclusion problem is decidable for all the measures:

\begin{thm}
  Let $V{\in}\{\Sum,\Avg,\Dsum,\Ratio\}$ and
  let $A,B$ be two $V$-automata where $B$ is  functional.
  The inclusion problem 
  $L_A{\leq L_B}$ is decidable. Moreover, if $V{\in}\{\Sum,\Avg,\Dsum\}$ then it is
  {\sf PSpace-c} and if additionally $B$ is deterministic, it is in
  \ptime.
\end{thm}
   
\begin{proof}
  Let $V\in\{\Sum,\Avg,\Dsum\}$. In a first step, we test the
  inclusion of the domains $\dom(A)\subseteq \dom(B)$ (it is well-known from theory of finite automata to be in {\sf
    PSpace-c} and in \ptime if $B$ is deterministic). Then 
   we construct the product $A\times B$ as follows:
   $(p,q)\xrightarrow{a|n_A-n_B} (p',q')\in\delta_{A\times B}$ iff
   $p\xrightarrow{a|n_A} p'\in\delta_A$ and $q\xrightarrow{a|n_B}
   q'\in\delta_B$. Then $L_A\not\leq L_B$ iff $L_{A\times B}^{>0}\neq \varnothing$, which
is decidable by Theorem \ref{teoEmpty}.


   Let $V=\Ratio$. As for the other measures we first check inclusion
   of the domains. We then define the product $A\times B$ of $A$ and $B$
   as a labelled transition system whose set of transitions
   $\delta_{A\times B}$ is defined by $(p,q)\xrightarrow{a|(r_1,c_1,r_2,c_2)} (p',q')\in\delta_{A\times B}$ iff
   $p\xrightarrow{a|(r_1,c_1)} p'\in\delta_A$ and $q\xrightarrow{a|(r_2,c_2)}
   q'\in\delta_B$. For all $t\in \delta_{A\times B}$, we let $r_A(t)$ be the reward 
   of the transition $t$ projected on $A$. The values $c_A(t), r_B(t)$ and 
   $c_B(t)$ are defined similarly. We let $\mathcal{P}(A\times B)$ be the
   Parikh image of the transitions of $A\times B$, i.e. the set of 
   total functions $\alpha:\delta_{A\times B}\rightarrow \mathbb{N}$ such that
   there exist $w\in\Sigma^+$ and a path labelled $w$ from the pair
   of initial states to a
   pair of accepting states, such that this path passes by $t$ exactly
   $\alpha(t)$ times, for all $t\in \delta_{A\times B}$.
   It is well-known by Parikh's theorem
   that $\mathcal{P}(A\times B)$ can be effectively represented
   by a set of linear constraints \cite{Parikh}.
   We now define the set of vectors $\Gamma$ that are the Parikh images
   of accepting runs of $A\times B$ which, when projected on $A$, has a strictly bigger
   ratio value than the one obtained by the projection on $B$.  In particular, the the set of vectors $\Gamma$ is given by:
   $$
    \{ \alpha:\delta_{A\times B}\rightarrow \mathbb{N}\ |\ \alpha\in \mathcal{P}(A\times B), \frac{\sum_{t\in\delta_{A\times B}}\alpha(t)\cdot r_A(t)}{\sum_{t\in\delta_{A\times B}}\alpha(t)\cdot c_A(t)} > \frac{\sum_{t\in\delta_{A\times B}}\alpha(t)\cdot r_B(t)}{\sum_{t\in\delta_{A\times B}}\alpha(t)\cdot c_B(t)} \} 
   $$
   It is easy to check that $\Gamma\neq\varnothing$ iff $L_A\not\leq L_B$. The set $\Gamma$ can be defined as the solutions over natural numbers of a system of equations in linear and  quadratic forms (i.e. in which products of two variables are
   permitted). There is one variable $x_t$ for each $t\in\delta_{A\times B}$ that 
   gives the number of times $t$ is fired in an accepting run of $A\times B$. It is decidable whether such a system has a solution
   \cite{KariantoKT06,Grunewald04}. 
\end{proof}

There is no known complexity bound for solving quadratic equations,
so the proof above does not give us a complexity bound for the inclusion problem of functional
\Ratio-automata. However, thanks to the functionality test, which is
in {\sf PSpace} for \Ratio-automata, we can test equivalence of two functional
\Ratio-automata $A_1$ and $A_2$ in {\sf PSpace}: 

\begin{thm}
  Let $V\in\{\Sum,\Avg,\Dsum,\Ratio\}$. Equivalence of functional
  $V$-automata is {\sf PSpace-c}.
\end{thm}
\begin{proof}
The following algorithm can be used to test equivalence for all the considered measures: first check in {\sf
  PSpace} that $\dom(A_1)=\dom(A_2)$ using the standard equivalence algorithm for non-deterministic finite automata. Then, check that the union of $A_1$
and $A_2$ is functional. The latter can be done in  {\sf CoNP} for \Ratio-automata and in polynomial time for the other measures, using the functionality tests defined in the previous section (cf. Theorem \ref{teocorsum}, Corollary \ref{corfunc}, and Theorem \ref{thm:funratio}).   
\end{proof}


\section{Realizability Problem}
In this section, we consider the problem of \emph{quantitative language
realizability}.  The
realizability problem is better understood as a
game between two players: the 'Player input' (the environment, also
called Player $I$)  and the 'Player output' (the controller, also
called Player $O$). Player $I$ (resp. Player $O$) controls
the letters of a finite alphabet $\Sigma_I$ (resp. $\Sigma_O$). We
assume that $\Sigma_O\cap \Sigma_I=\varnothing$ and that $\Sigma_O$
contains the special symbol $\dashv$ whose role is to stop the game. We
let $\Sigma = \Sigma_O\cup \Sigma_I$.

Formally, the realizability game is   a turn-based game played on an arena defined by a weighted automaton  
$A {=} (Q{=}Q_O\uplus Q_I,
q_I,F,\delta,\gamma)$,  whose set of states is
partitioned into two sets $Q_O$ and $Q_I$,  and such that $F\subseteq
Q_I$ and $\delta{\subseteq} (Q_O{\times} \Sigma_O{\times} Q_I)\cup
(Q_I{\times} \Sigma_I{\times} Q_O)$. Recall that by definition of the
transitions of weighted automaton $\dom(A){\subseteq} (\Sigma\backslash \{\dashv\})^*\dashv$. Player $I$
starts by giving an initial letter $i_0\in\Sigma_I$, 
Player $O$ responds providing a letter $o_0\in\Sigma_O$, then Player $I$ gives $i_1$ and
Player $O$ responds $o_1$, and so on. Player $O$ has also the power to
stop the game at any turn with the distinguishing symbol $\dashv$. In this
case, the game results in a finite word $(i_0o_0)(i_1o_1)\dots
(i_j\dashv)\in \talpha$, otherwise the outcome of the game is an infinite
word  $(i_0o_0)(i_1o_1)\dots \in (\Sigma\setminus\{\dashv\})^\omega$.

The players play according to strategies. A strategy for Player $O$ (resp. Player $I$) is a
mapping $\lambda_O: (\Sigma_I\Sigma_O)^*\Sigma_I\rightarrow \Sigma_O$
(resp. $\lambda_I: (\Sigma_I\Sigma_O)^*\rightarrow \Sigma_I$). The
outcome of the strategies $\lambda_O,\lambda_I$ is the word
$w=i_0o_0i_1o_1\dots$ denoted by $\outcome(\lambda_I,\lambda_O)$ such
that for all $0\leq j\leq |w|$ (where $|w| = +\infty$ if $w$ is
infinite), $o_j = \lambda_O(i_0\dots i_{j-1})$ and
$i_j=\lambda(i_0\dots o_j)$, and such that if $o_j = {\dashv}$ for some
$j$, then $w=i_0\dots o_j$. We denote by $\Lambda_O$ (resp. $\Lambda_I$) the set of strategies for
Player $O$ (resp. Player $I$). 
A strategy $\lambda_O\in\Lambda_O$ is winning for Player $O$ if for all 
$\lambda_I\in \Lambda_I$, $\outcome(\lambda_I,\lambda_O)$
is finite and $L_A(\outcome(\lambda_I,\lambda_O))> \nu$, where $\nu\geq 0$ a given threshold. 

Given  a weighted automaton $A$ and a threshold $\nu\geq 0$, the \textit{quantitative language
realizability problem} 
 asks whether Player $O$ has a winning strategy and in that case, we
 say that $A$ is \textit{realizable}.
Our first result on realizability  is 
 negative: in particular, Subsection \ref{sub:undecReaz}   shows that the realizability problem is
 undecidable for weighted functional \Sum-, \Avg-automata, and \Ratio-automata. However, when \emph{deterministic} weighted automata are considered, we can provide positive decidability results for all the considered measures (cfr. subsection \ref{sub:decReaz}).

 \subsection{Undecidability Results}\label{sub:undecReaz} 
  We show that  the halting problem for deterministic
 $2$-counter Minsky machines ~\cite{minsky67finite} can be reduced to the quantitative
 language realizability problem for (functional) \Sum-automata (resp. \Avg-automata). This entails our undecidability results w.r.t. realizability for $V$-automata ($V\in \{\mbox{\Sum,\Avg,\Ratio}\}$).
 
\subsubsection*{$2$-Counter Machines} A \emph{$2$-counter machine}~$M$ consists
of a finite set of control states $S$, an initial state $s_I \in S$, a final state $s_F \in Q$,
a set $C$ of counters ($\abs{C} = 2$) and a finite set $\delta_M$ of instructions manipulating
two integer-valued counters. Instructions are of the form
\begin{itemize}[label=$s:$]
\item[$s:$] $c := c + 1$ {\bf goto } $s'$
\item[$s:$] {\bf if } $c = 0$  {\bf then goto } $s'$  {\bf else }$c := c - 1$ {\bf goto $s''$}.
\end{itemize}
Formally, instructions are tuples $(s, \alpha, c, s')$  where $s,s' \in S$ are source and target states respectively,
and the action $\alpha \in \{inc, dec, 0?\}$ applies to the counter $c \in C$. We assume that~$M$
is deterministic: for every state $s \in S$, either there is exactly one instruction $(s, \alpha,\cdot,\cdot) \in \delta_M$
and $\alpha = inc$, or there are two instructions $(s, dec, c, \cdot), (s, 0?, c, \cdot) \in \delta_M$.

A \emph{configuration} of~$M$ is a pair $(s, v)$
where $s \in S$ and $v: C \to \nat$ is a valuation of the counters.
An \emph{accepting run} of~$M$ is a finite sequence $\pi = (s_0,v_0) \delta_0 (s_1,v_1) \delta_1  \dots \delta_{n-1} (s_n,v_n)$
where $\delta_i = (s_i, \alpha_i, c_i, s_{i+1}) \in \delta_M$ are instructions and  $(s_i,v_i)$ are configurations of~$M$ such that
$s_0 = s_I$, $v_0(c) = 0$ for all $c \in C$, $s_n = s_F$, and for all $0 \leq i < n$, we have $v_{i+1}(c) = v_i(c)$ for $c \neq c_i$,
and
$(a)$ if $\alpha = inc$, then $v_{i+1}(c_i) = v_i(c_i) + 1$
$(b)$ if $\alpha = dec$, then $v_i(c_i) \neq 0$ and $v_{i+1}(c_i) = v_i(c_i) - 1$, and
$(c)$ if $\alpha = 0?$, then $v_{i+1}(c_i) = v_i(c_i) = 0$.
The corresponding \emph{run trace} of $\pi$ is the sequence of instructions $\bar{\pi} = \delta_0 \delta_1 \dots \delta_{n-1}$.
The \emph{halting problem} is to decide, given a 2-counter machine~$M$, whether~$M$ has an accepting run.
This problem is undecidable~\cite{minsky67finite}.

\subsubsection*{The encoding}  We are now ready to present our encoding. The latter goes along the lines of the encodings in \cite{DDGRT10,AlmagorBokerKupferman2011} to prove undecidability results related to imperfect information games and weighted automata. 

Given a deterministic $2$-counter machine $M$, we construct a functional  \Sum-automaton $A=(Q,q_I,F,\delta,\gamma)$ (resp. \Avg-automata), where $Q=Q_O\cup Q_I\cup F$, $\Sigma=\Sigma_O\cup\Sigma_I$  and $\delta\subseteq Q\times \Sigma\times Q$, such that $M$ halts if and only if $L(A)$ is realizable (with realizability threshold $\nu=0$). In particular, $\Sigma_O=\delta_M$ and
a strategy $\lambda_O\in \Lambda_O$ for Player~$O$ is winning if and only if    for each $\lambda_I\in \Lambda_I$, the projection of $\outcome(\lambda_I,\lambda_O)$ onto $\Sigma_O$ is  an accepting run of $M$. The alphabet $\Sigma_I$  for Player~$I$ is the set of letters $\Sigma_I=\{go\}\cup (\bigcup_{i=1,2}\{\emph{cheatCi+},  \emph{cheatCi-}\}) \cup (\bigcup_{0\leq j<|S|}\{\emph{cheatR:$s_j$} \})$. 

Intuitively, the role of Player~$I$ is that of observing the play of Player~$O$ and detecting whether he faithfully simulates $M$, or he cheats.
If Player~$O$ cheats by declaring the $i$-th counter equal to $0$  when it is not (positive cheat), then Player~$I$ can use the action
\emph{cheatCi+}, $i\in \{1,2\}$, to force all the runs but one (with  weight $\leq 0$) to die.
Similarly, if Player~$O$ cheats by decrementing a counter with value zero (negative cheat) or on the structural properties of a run of $M$, then Player~$I$ can win by playing the corresponding observing action : \emph{cheatCi-}, for negative cheats on counter $i\in\{1,2\}$, or  $\emph{cheatR:$s_j$}$ for a cheat on the run through $M$ detected at state $s_j$.

In detail, the automaton $A$ consists of an initial state $q_I$ from which Player~$I$ can nondeterministically jump to several gadgets: each gadget checks one of
the properties of the sequence of actions provided by Player~$O$, and
verify whether Player~$O$ simulates faithfully $M$ or he eventually cheats.  More specifically, $q_I$ has one transition with weight $0$ and label \emph{go} to the set of gadgets listed below and described in detail in the rest of this subsection:
\begin{itemize}
\item two gadgets to observe positive cheats over a counter (one gadget for each counter $i\in\{1,2\}$);
\item  two gadgets to observe negative cheats over a counter (one gadget  for each counter $i\in\{1,2\}$);
\item a gadget to observe a structural cheat for each state $s\in S$ that can be traversed by a path in $M$;
\item a neutral gadget, where
Player~$I$ simply observes the run provided by Player~$O$ and let such a run to reach a final state as soon as Player~$O$ provides an action simulating a step toward the halting state of $M$.
\end{itemize}

\noindent Due to the initial nondeterministic choice, each final state (in one of the gadget) is accessible throughout the evolution of the play and Player~$O$ has to ensure that all the properties checked in the gadgets are fulfilled. Otherwise, Player~$I$
will have the ability to use a letter in $\Sigma_I\setminus\{go\}$ to  
 let just one run (in the appropriate gadget)  to  survive, ensuring that such a run eventually   reaches the final state with weight $\leq 0$.

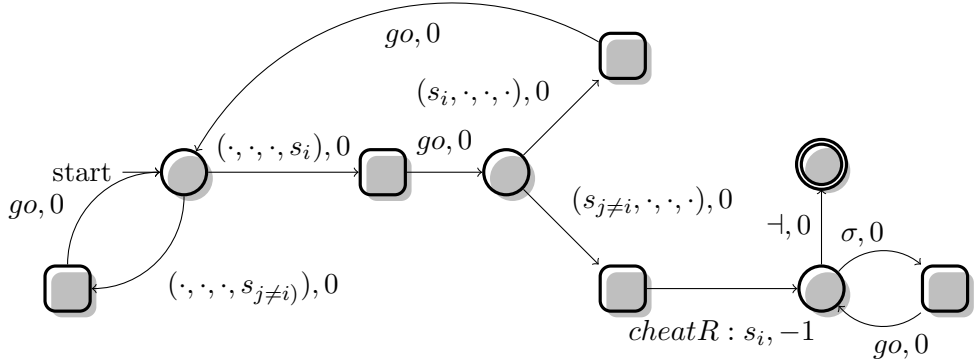
\begin{figure}
\hrule
  \begin{center}
\def\fsize{\normalsize}

\begin{tikzpicture}


\tikzset{nodeP2/.style={rectangle,draw,very thick,minimum size=6mm, rounded corners,drop shadow}}
\tikzset{nodeP1/.style={circle,draw,very thick,minimum size=6mm,drop shadow}}
\tikzset{nodeP1init/.style={initial,circle,draw,very thick,minimum size=6mm,drop shadow}}
\tikzset{nodeP1final/.style={accepting,circle,draw,very thick,minimum size=6mm,drop shadow}}
\node[nodeP1init](x1) {};
\node[nodeP2](x2) [right=2cm of x1] {};
\node[nodeP2](x3) [below left=of x1] {};
\node[nodeP1](x4) [right=of x2] {};
\node[nodeP2](x5) [above right=of x4] {};
\node[nodeP2](x6) [below right=of x4] {};
\node[nodeP1](x7) [right=2cm of x6] {};
\node[nodeP2](x8) [right=of x7] {};
\node[nodeP1final](x9) [above=of x7] {};

\draw [->] (x1) to node [auto,midway] {$(\cdot,\cdot,\cdot,s_i),0$} (x2);
\draw [->] (x1) to [bend left=45] node [midway,auto] {$(\cdot,\cdot,\cdot,s_{j\neq i)}),0$} (x3);
\draw [->] (x3) to [bend left=45] node [near start,auto] {$go,0$} (x1);
\draw [->] (x2) to [] node [midway, yshift=4mm] {$go,0$} (x4);
\draw [->] (x4) to node [auto,midway] {$(s_i,\cdot,\cdot,\cdot),0$} (x5);
\draw [->] (x4) to node [auto,midway] {$(s_{j\neq i},\cdot,\cdot,\cdot),0$} (x6);
\draw [->] (x5) to [bend right=45] node [auto, midway] {$go, 0$} (x1);
\draw [->] (x6) to node [auto,midway, yshift=-9mm] {$cheatR:s_i,-1$} (x7);
\draw [->] (x7) to [bend left=45] node [auto, near start, xshift=5mm] {$\sigma, 0$} (x8);
\draw [->] (x8) to [bend left=45] node [auto, near start,  xshift=-5mm] {$go, 0$} (x7);

\draw [->] (x7) to  node [auto] {${\dashv}, 0$} (x9);

\end{tikzpicture}
  \end{center}
\hrule
\caption{Gadget to check cheats along the run. \label{fig:gadget5Func1}}
\end{figure}

\begin{figure}
\hrule
  \begin{center}
\def\fsize{\normalsize}

\begin{tikzpicture}


\tikzset{nodeP2/.style={rectangle,draw,very thick,minimum size=6mm, rounded corners,drop shadow}}
\tikzset{nodeP1/.style={circle,draw,very thick,minimum size=6mm,drop shadow}}
\tikzset{nodeP1init/.style={initial,circle,draw,very thick,minimum size=6mm,drop shadow}}
\tikzset{nodeP1final/.style={accepting,circle,draw,very thick,minimum size=6mm,drop shadow}}
\node[nodeP1init](x1) {};
\node[nodeP2](x2) [right=of x1] {};
\node[nodeP2](x3) [right=of x2] {};
\node[nodeP1](x4) [right=of x3] {};
\node[nodeP1final](x5) [right=of x4] {};

\draw [->] (x1) to node [auto,midway] {$\sigma,0$} (x2);
\draw [->] (x2) to [bend left=45] node [auto,  xshift=10mm] {$go,0$} (x1);
\draw [->] (x1) to [bend left=45] node [auto,  midway] {$(\cdot,\cdot,\cdot,s_{F}), 0$} (x3);
\draw [->] (x3) to [auto, swap,  midway] node  {$go, 0$} (x4);
\draw [->] (x4) to [auto] node  {${\dashv}, 1$} (x5);


\end{tikzpicture}
  \end{center}
\hrule
\caption{Neutral gadget. \label{fig:gadget3Func1}}
\end{figure}
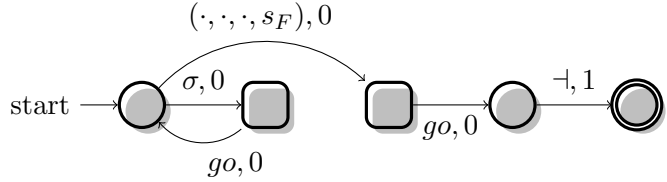

We are now ready to present in detail the  gadgets (cf. Figure \ref{fig:gadget1Func1}--\ref{fig:gadget3Func1}). In particular,  in each gadget the states  belonging to $Q_I$ (resp. $Q_O$) are represented by a square node (resp. circle node), while final states are double lined\footnote{The entering node (owned by Player~$I$, with an incoming edge labeled \emph{start}) is supposed to be connected to $q_I$ (the initial state of $A$, owned by Player~$I$) by means of a transition  on the symbol \emph{go}$\in\Sigma_I$ with weight $0$}. Each transition in the gadgets leaving a node owned by Player~$I$ is labelled by a pair  of the form $(\sigma,w)$, where $\sigma\in\Sigma_I$ and $w\in\{-1,0,1\}$ is the weight of the transition. Symmetrically, each transition in the gadgets leaving a circle-node (i.e. a node owned by Player~$O$) is labelled by a pair of the form  $(\sigma,w)$, where $w\in\{-1,0,1\}$ is the weight and $\sigma\in\delta_M$ is an instruction of the counter machine. Namely, $\sigma$ is of the form $(s,\alpha,i,s')$, where $i\in\{1,2\}$ represents the counter, $\alpha \in \{inc, dec, 0?\}$ and $s$ (resp. $s'$) is the sourcing (resp. target) state. We use the notation $\cdot$ when a symbol within an instruction could be any admissible (e.g. $(\cdot,inc,1,\cdot)$ represents any instruction incrementing the first counter). Finally $\sigma$ denotes an arbitrary letter in the alphabeth of the player.

\begin{figure}
\begin{minipage}{.5\textwidth}
\hrule
\begin{center}
\def\fsize{\normalsize}

\begin{tikzpicture}


\tikzset{nodeP2/.style={rectangle,draw,very thick,minimum size=6mm, rounded corners,drop shadow}}
\tikzset{nodeP1/.style={circle,draw,very thick,minimum size=6mm,drop shadow}}
\tikzset{nodeP1init/.style={initial,circle,draw,very thick,minimum size=6mm,drop shadow}}
\tikzset{nodeP1final/.style={accepting,circle,draw,very thick,minimum size=6mm,drop shadow}}

\node[nodeP1init](x1) {};
\node[nodeP2](x2) [right=of x1] {};
\node[nodeP2](x3) [below=of x1] {};
\node[nodeP1](x4) [below=of x3] {};
\node[nodeP2](x5) [right=of x4] {};
\node[nodeP1final](x6) [left=of x4] {};

\draw [->] (x1) to [bend left=45] node [auto] {$\begin{array}{lr}
(\cdot, inc, i, \cdot), & \minusone \\
(\cdot, dec, i, \cdot), & 1 \\
\end{array}$} (x2);

\draw [->] (x2) to [bend left=45] node [auto, near start] {$go, 0$} (x1);
\draw [->] (x3) to [bend right, midway, auto, swap, near start] node  {$go, 0$} (x1);
\draw [->] (x1) to [auto, swap, bend right,  midway] node  {$(\cdot, 0?, i, \cdot), 0$} (x3);
\draw [->] (x3) to  node [auto,swap] {$cheatCi+, 0$} (x4);
\draw [->] (x4) to [bend left=45] node [auto, near end] {$\sigma, 0$} (x5);

\draw [->] (x5) to [bend left=45] node [auto, near end] {$go, 0$} (x4);
\draw [->] (x4) to node [auto] {${\dashv}, 1$} (x6);


\end{tikzpicture}
\end{center}
\hrule
 \caption{Gadget to check positive cheats\label{fig:gadget1Func1}}
\end{minipage}
\quad
\begin{minipage}{.5\textwidth}
\hrule
\begin{center}
\def\fsize{\normalsize}

\begin{tikzpicture}


\tikzset{nodeP2/.style={rectangle,draw,very thick,minimum size=6mm, rounded corners,drop shadow}}
\tikzset{nodeP1/.style={circle,draw,very thick,minimum size=6mm,drop shadow}}
\tikzset{nodeP1init/.style={initial,circle,draw,very thick,minimum size=6mm,drop shadow}}
\tikzset{nodeP1final/.style={accepting,circle,draw,very thick,minimum size=6mm,drop shadow}}
\node[nodeP1init](x1) {};
\node[nodeP2](x2) [right=of x1] {};
\node[nodeP2](x3) [below=of x1] {};
\node[nodeP1](x4) [below=of x3] {};
\node[nodeP2](x5) [right=of x4] {};
\node[nodeP1final](x6) [left=of x4] {};
\draw [->] (x1) to [bend left=45] node [auto] {$\begin{array}{lr}
(\cdot, 0?, i, \cdot), & 0\\
(\cdot, inc, i, \cdot), & 1 \\
\end{array}$} (x2);

\draw [->] (x2) to [bend left=45] node [auto,  near start] {$go, 0$} (x1);
\draw [->] (x3) to [auto, swap, bend right, near  start] node  {$go, 0$} (x1);
\draw [->] (x1) to [auto, swap, bend right,  midway] node  {$(\cdot, dec, i, \cdot),  \minusone$} (x3);
\draw [->] (x3) to  node [auto,swap] {$cheatCi-, 0$} (x4);
\draw [->] (x4) to [bend left=45] node [auto, near end] {$\sigma, 0$} (x5);

\draw [->] (x5) to [bend left=45] node [auto, near end] {$go, 0$} (x4);
\draw [->] (x4) to node [auto] {${\dashv}, 1$} (x6);

\end{tikzpicture}
\end{center}
\hrule
 \caption{Gadget to check negative cheats\label{fig:gadget2Func1}}
\end{minipage}
\end{figure}
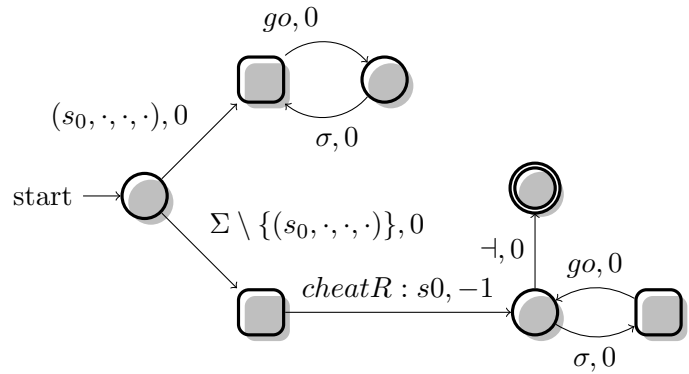
\begin{figure}
\hrule
  \begin{center}
\def\fsize{\normalsize}

\begin{tikzpicture}


\tikzset{nodeP2/.style={rectangle,draw,very thick,minimum size=6mm, rounded corners,drop shadow}}
\tikzset{nodeP1/.style={circle,draw,very thick,minimum size=6mm,drop shadow}}
\tikzset{nodeP1init/.style={initial,circle,draw,very thick,minimum size=6mm,drop shadow}}
\tikzset{nodeP1final/.style={accepting,circle,draw,very thick,minimum size=6mm,drop shadow}}
\node[nodeP1init](x1) {};
\node[nodeP2](x2) [above right=of x1] {};
\node[nodeP2](x3) [below right=of x1] {};
\node[nodeP1](x4) [right=of x2] {};
\node[nodeP1](x5) [right=3cm of x3] {};
\node[nodeP2](x6) [right=of x5] {};
\node[nodeP1final](x7) [above=of x5] {};

\draw [->] (x1) to node [auto,midway] {$(s_0,\cdot,\cdot,\cdot),0$} (x2);
\draw [->] (x2) to [bend left=45] node [auto,xshift=-10mm] {$go,0$} (x4);
\draw [->] (x4) to [bend left=45] node [auto,xshift=-3mm] {$\sigma,0$} (x2);
\draw [->] (x1) to node [auto,midway] {$\Sigma\setminus\{(s_0,\cdot,\cdot,\cdot)\},0$} (x3);
\draw [->] (x3) to  node [auto,midway] {$cheatR:s0, -1$} (x5);
\draw [->] (x5) to [auto, swap, bend right, midway] node  {$\sigma, 0$} (x6);
\draw [->] (x6) to [auto, swap, bend right, midway] node  {$go, 0$} (x5);
\draw [->] (x5) to [auto] node  {${\dashv}, 0$} (x7);

\end{tikzpicture}
  \end{center}
\hrule
\caption{Gadget to check that Player $O$ plays $(s_0,\cdot,\cdot,\cdot)$ at the beginning. \label{fig:gadget4Func1}}
\end{figure}

Given the above notation, Figure \ref{fig:gadget1Func1}  represents the gadget to check a positive cheat on counter $i$, $i\in\{1,2\}$. Player~$I$ observes the inverted value of the counter $i$ throughout the path on $M$ simulated by Player~$O$. Whenever Player~$O$ declares that counter $i$ is equal to $0$, Player~$I$ can use the action \emph{cheatCi+} to kill all the runs in $A$ but the one within the observing gadget. The evolution of  such a run up to  \emph{cheatCi+}  will have a negative value (corresponding to the inverted value of the observed counter) if  Player~$O$ was cheating. Hence, as soon as Player $O$ playes $\dashv$ it will end in a final state with weight $\leq 0$.
Symmetrically, the gadget for checking negative cheats (represented in Figure \ref{fig:gadget2Func1})  uses the weights on the edges to store the value of the observed counter. If  Player~$O$ cheats decrementing counter $i$ when its value is $0$,  Player~$I$ can use the action \emph{cheatCi-} to kill all the runs but the one (with negative value) in the gadget observing negative cheats.

Player~$I$ can use the gadgets in Figures \ref{fig:gadget4Func1}--\ref{fig:gadget5Func1}   to detect any structural cheat committed by Player~$O$. If Player~$O$ initially provides an  action different from $(s_0,\_,\_,\_)$, Player~$I$ can punish him by playing action \emph{cheatR:s0}. Similarly, if Player~$O$ provides two actions that do not induce a (sub)-path in $M$, Player~$I$  can punish him within the gadget in Figure \ref{fig:gadget5Func1}. Finally, Figure \ref{fig:gadget3Func1} illustrates the neutral gadget, where
Player~$I$ simply observes the run provided by Player~$O$ and let such a run to reach a final state as soon as Player~$O$ provides an action simulating a step toward the halting state of $M$.

\begin{thm}\label{realizabilityUndec}
Let $V\in\{\Sum,\Avg,\Ratio\}$. The realizability problem for
functional $V$-automata is undecidable.
\end{thm}
\begin{proof}
 We show that the encoding outlined above is correct  by proving that $M$ halts iff  Player~$O$ has a strategy to win the realizability game on the \Sum-automaton (resp. \Avg-automaton) $A$. Namely, we show that Player~$O$ wins the realizability game iff he provides a word $\pi$ which corresponds to an accepting run of $M$ (and then stop the game). 

\noindent ($\Rightarrow$) Suppose that $M$ halts. Let $\pi$ be the run of $M$ leading to the halting state,  and consider  $\lambda_O(\pi)\in\Lambda_O$, where $\lambda_O(\pi)$ denotes the strategy for Player~$O$ induced by $\pi$, in which Player$O$ provides the word $\pi$ and then stop the game.
Let $\lambda_I\in \Lambda_I$. There are two cases to consider.
\begin{enumerate}
\item In the first case, $\lambda_I$ does not provide any action in: $$(\bigcup_{i=1,2}\{\emph{cheatCi+},  \emph{cheatCi-}\}) \cup (\bigcup_{0\leq j<|S|}\{\emph{cheatR:$s_j$} \})$$ 
Then, the only run to a final state in $A$ is the one within the neutral gadget, having weight strictly positive.
\item In the second case, $\gamma_I$  contains an action  in $(\bigcup_{i=1,2}\{\emph{cheatCi+},  \emph{cheatCi-}\})$. Let $\alpha$ be the first action in $(\bigcup_{i=1,2}\{\emph{cheatCi+},  \emph{cheatCi-}\}) $ on $\lambda_I$.
    There is only one gadget allowing a run containing $\alpha$. Since $\pi$ is faithfully simulating $M$, such a run  leads to a final state in the corresponding gadget with value $>0$.

 \end{enumerate}
 Note that $\lambda_I$ cannot contain an action  $\alpha\in  (\bigcup_{0\leq j<|S|}\{\emph{cheatR:$s_j$} \})$. In fact,  Player~$I$ can never play $\emph{cheatR:$s_j$}$, since Player~$O$ does not commit any structural cheat on the run $\pi$.
 Hence, we conclude that $\forall \lambda_I\in \Lambda_I(L_A(\outcome(\lambda_O(\pi),\lambda_I)> 0))$.

 \noindent ($\Leftarrow$) Suppose that  $\lambda_O\in\Lambda_O$ is such that $\forall \lambda_I\in \Lambda_I(L_A(\outcome(\lambda_O,\lambda_I)\geq 0))$. By construction of $A$, $\lambda_O$   allows Player~$O$ to survive in  the gadgets for detecting positive, negative or structural cheats if and only if the projection of the outcome onto $\Sigma_O$ is a faithful simulation of a run in $M$. If Player~$I$ can not use an action in $(\bigcup_{i=1,2}\{\emph{cheatCi+},  \emph{cheatCi-}\}) \cup (\bigcup_{0\leq j<|S|}\{\emph{cheatR:$s_j$} \})$ to win (using the  gadget targeted to check the corresponding cheat), the only remaining strategy for Player~$I$ is playing indefinitely $\neg cheat$. In that case, Player~$O$ wins only if he eventually provides an action simulating a step leading to an halting state in $M$ (and then stop the game). Thus, our hypothesis entail that $\lambda_O$ consists in providing a run for $M$ that leads to a final state, witnessing  that $M$ halts.
 Undecidability for \Ratio-automata follows from the fact that any \Avg-automaton  $A = (Q,q_I,F,\delta,\gamma)$ can be coded into a \Ratio-automaton by reweighting from $\gamma(t)$ to $(\gamma(t),1)$ each transition $t\in\delta$.
\end{proof} 

\subsection{Realizability on Deterministic Weighted Automata}\label{sub:decReaz}
The encoding in the previous subsection  relies on the
use of a nondeterminism. When the automata are deterministic,
we recover decidability by considering suitable variants of 
classical games played on graphs, and prove that they are solvable in
${\sf NP}\cap{\sf coNP}$.

\begin{thm}\label{thm:gamedeter}
Let $V\in\{\Sum,\Avg,\Dsum,\Ratio\}$. The realizability problem for
deterministic $V$-automata is in ${\sf NP}\cap{\sf coNP}.$
\end{thm}
\begin{proof}
We first consider the case of deterministic \Sum-automata. Let $A =
(Q=Q_O\uplus Q_I,q_I,F,\delta,\gamma)$ be a
deterministic \Sum-automaton. Without loss of generality, we assume
that $A$ contains only one accepting state denoted by $f$ which is
absorbing\footnote{Note that in our setting Player O has the ability
  to stop the game by means of the special symbol $\dashv$. Therefore,
  it is sufficient to redirect the transition labeled by $\dashv$ to
  a unique absorbing final state.}. Then we consider $A$ as a finite
state game arena and compute the set of states $S \subseteq Q$ from
which player~$O$ can force a visit to the accepting state $f$. Note
that from any state $s$ in $S$, player~$O$ has a strategy to force a
visit to $f$ within $n$ steps, where $n=|Q|$. Note also that by
determinacy, the complement of this set is the set of states of $A$
from which player~$I$ has a strategy to prevent a visit to
$f$. Clearly, player~$O$ has to avoid the states in $Q \setminus S$ at
all cost and so they can be removed from $A$. Let $A'$ be $A$ where we
have kept only the states in $S$.

Now, we construct from $A'$ a finite tree as follows. We unfold $A'$ and stop a branch at a node when:
  \begin{itemize}
  	\item it is labeled with $f$ and the sum of the weights on the branch up to the node is equal to $c > \nu$,
	\item it is labeled by a state $q$ that already appears on the branch from the root to the node. We call the node where $q$ already appears the {\em ancestor} of the leaf.
  \end{itemize}
\noindent
Let us note $L$ the set of leafs of this finite tree. We then partition the leafs of this tree into $L_1$, the set of leafs that are good for player~$O$ and $L_2$, the set of leafs that are good for player~$I$. $L_1$  contains:
  \begin{itemize}
  	\item $(C_1)$ the leafs that are annotated with $f$ and for which the sum of weights is strictly greater than the realizability threshold $\nu$, and 
	\item $(C_2)$ the leafs labeled with a repeating state and for which the sum of weights from the root to the leaf is strictly larger than the sum of weights from the root to the ancestor.   
  \end{itemize}
 \noindent
 $L_2=Q \setminus L_1$ are the leafs that are good for player~$I$. Now, consider the game played on this finite tree where player~$O$ wants to reach $L_1$ and player~$I$ wants to reach $L_2$. The winner in this game can be found by backward induction. We claim (and prove) below that player~$I$ wins in this finite game tree iff he wins the original game.
 
Assume that player~$O$ wins the finite game tree. We show how to construct a winning strategy in the original game. The strategy is built as follows. In the original game, player~$O$ plays as in the final tree up to the point where he reaches a leaf (in $L_1$). If the leaf is of sort defined in $C_1$ above then we know that player~$O$ has won the original game. Otherwise, we know that the sum now is strictly greater than the sum up to the ancestor of the leaf that we have reached. Then player~$O$ continues to play as prescribed by its winning strategy in the tree from the ancestor. Continuing like that, each time that the game arrives at a leaf, the sum of weights has strictly increased from the last visit to that leaf. As a consequence, after a finite amount of time, the sum will be strictly larger than $\nu+n \cdot |-W|$ where $-W$ is the smallest negative weight in $A'$. From that point, player~$O$ can use his strategy that forces the state $f$ and reach it with a sum that is strictly greater than $\nu$  (this is because he can force $f$ within $n$ steps).
 
Now assume that player~$I$ wins the finite game tree. We show how to construct a winning strategy in the original game. The strategy simply follows the strategy of player~$I$ in the finite tree by applying the strategy from the ancestor when reaching a leaf. As only leaf in $L_2$ are reached when playing that way, we know that the sum on successive visits to repeating states is non-increasing. As a consequence, as player~$O$ can not force a visit to a node labeled with $f$ and  sum greater than $\nu$ in the finite game tree, we know that this will not happen in the original game neither when player~$I$ plays its strategy.

This proof establishes that the realizability problem is decidable for deterministic \Sum-automata. Note that player~$O$ needs memory to win in the original game as he has to verify that he has reached a sufficiently high sum before applying the strategy that forces the visit to $f$. To provide the complexity bound of ${\sf NP}\cap{\sf coNP}$, we reduce the realizability problem for deterministic \Sum-automata to the decision problem on mean-payoff games \cite{EM79,ZP96}. A mean-payoff game  is a  zero-sum game involving two players (the maximazer $P_0$ and the minimizer $P_1$) where $P_0$ has the following objective: to maximize  his mean-payoff, defined as the liminf of the  ratio between the cost of a play-prefix of length $l$ (i.e. the sum of the weights along the path) and $l$ when $l\rightarrow +\infty$.

  Given a deterministic \Sum-automaton $A = (Q=Q_O\uplus Q_I,q_I,F,\delta,\gamma)$, where $f$ is an absorbing accepting state, consider the mean-payoff game $G=(V_0,V_1,\delta_G,\gamma)$, where $V_0\subseteq Q_O$ (resp. $V_1\subseteq Q_I$) is the set of states of player $O$ (resp. player $I$) for which player $O$ has a strategy to force a visit to $f$. The transition relation $\delta_G$ is built as follows. We add an edge from $f$ back to $q_I$ (weighted $0$)  and we omit the self-loop on the absorbing final state.  Moreover, if $\nu>0$ then $\delta_G(q,q')=\delta(q,q')-\nu$ for each edge $(q,q')$ in $A$. We prove that player $O$ has a winning strategy w.r.t. realizability on $A$ if and only if $val(q_I)>0$ in the mean-payoff game $G$. Suppose $val(q_I)>0$ in $G$. Then, player $O$ has a memoryless strategy to ensure $val(q_I)>0$. Let $\sigma$ be such a memoryless strategy for player $O$, and consider the graph $G_\sigma$ obtained from $G$ by removing all the edges sourcing from a vertex in $V_0$ that are not chosen according to $\sigma$. Since $\sigma$ guarantees $val(q_I)>0$, $G_\sigma$ does not contain any non-positive cycle. By construction, each simple cycle involving $f$ in $G$ is a simple path to $f$ followed by an edge back to $q_I$ with weight $0$. Hence, $G_\sigma$ contains a cycle through $f$ if and only if $G$ contains a simple positive path to $f$. Therefore, playing according to $\sigma$ on $A$, player $O$ will either reach $f$ with sum greater than $\nu$, or he will eventually reach a sum high enough to guarantee him to win, as soon as he forces the play to reach $f$. Conversely, assume that player $O$ has a strategy to win the realizability game on $A$ and let $\sigma$ be a corresponding winning strategy. To win the mean-payoff game $G$, player $O$ needs simply to apply $\sigma$ on $G$, since this leads him to $f$ with a positive sum, and then the play starts back from $q_I$ following the only edge sourcing from $f$ back to $q_I$.
As shown in  the previous results \Avg games can be reduced easily to \Sum games, and as for the questions about thresholds, \Ratio games can be reduced to \Avg games.

We now turn to the case of \Dsum. The solution for \Dsum is obtained
by first removing from $A$ all states from which player~$O$ cannot
force a visit to $f$. As above, we note the game where those states
have been removed by $A'$. Then, we consider $A'$ as an (infinite)
discounted sum game where player~$O$ tries to maximize the value of
the discounted sum while player~$I$ tries to minimize this value. Let
$v$ denote the value of the initial state $q_I$ in that game. We
claim that player~$O$ wins the initial game iff the value $v$ in $q_I$
is such that $v>\nu$. Indeed, if player~$O$ has a winning strategy in
the original game, i.e. a strategy to force the game into $f$ with
 discounted sum strictly greater than $\nu$, then by playing this strategy in the
discounted sum game, the infinite discounted sum will be equal to the
discounted sum up to $f$ as from there only the self loop on $f$ is
traversed and its weight is equal to $0$. Now assume that player~$O$ has
a strategy that force a value $v >\nu$ in the discounted sum game. Then
by playing that strategy for $i$ steps in the original game with $i$ large enough to make
 sure that $\lambda^iW+\dots +\lambda^{i+n}W$ is small enough, 
he will be able to switch to its strategy that forces
$f$ after at most $n$ steps and ensure to reach $f$ with a
 discounted sum $>\nu$. As infinite discounted sum games are
in ${\sf NP} \cap {\sf coNP}$ \cite{andersson2006} and since our reduction is polynomial, 
we also get that finite reachability discounted sum games are in
${\sf NP} \cap {\sf coNP}$. 
\end{proof}

\section{Determinizability Problem}
A weighted $V$-automaton $A$ is said to be \emph{determinizable} it
there exists a deterministic $V$-automaton $D$ such that $L_A(w) =
L_D(w)$ for all $w\in\talpha$. Weighted automata are not
determinizable in general. For example, 
consider the right automaton on Fig. \ref{fig:sumautomata}. Seen as
 \Sum-automaton, it cannot be determinized,
  because there are infinitely many delays associated with the pair
  of states $(p,q)$. Those delays can for instance be obtained 
by the family of words of the form $a^n$. Determinizability is already known to be decidable in \ptime for
functional \Sum-automata \cite{KirstenM05}\footnote{See
  \cite{DBLP:journals/tcs/KlimannLMP04,journals/mst/Kirsten06} for
  determinizability results on more general classes
  of \Sum-automata.}. It is also known that any \Dsum-automaton (even
non-functional) with an integral discount factor, i.e. $\lambda =
1/n$, where $n>1$ is an integer, is determinizable
\cite{DBLP:conf/csl/BokerH11}.

Determinizable functional $\Sum$-automata are characterized by the 
so called \emph{twinning property}, that has been introduced for
finite word transducers \cite{Choffrut77,BealEtAl03a}. In particular, the twinning property has been used as a sufficient condition
for the termination of Mohri's determinization algorithm \cite{Droste_Kuich_Vogler_2009} for
(non-functional) weighted automata over the tropical semiring. In \cite{KirstenM05}, Kirsten et al. proved that such a   property is also a necessary condition for determinization of functional \Sum-automata.  The twinning property has been also used  as a sufficient
condition for the termination of a determinization procedure, applied to classes of weighted automata defined on more general
commutative semirings \cite{KirstenM05} (the commutativity hypothesis is
necessary here to ensure that the twinning property is
sufficient). In this paper, we consider determinization of functional
arbitrary group automata. Further, we show that the twinning property is also a \emph{necessary}
condition for the determinization of functional group automata over groups
enjoying the so called infinitary property. We show that the groups
encoding \Sum, \Avg and \Dsum are all infinitary. Therefore, our general decidability result for the determinization
problem on functional infinitary group automata applies to functional\ \Sum, \Avg and \Dsum-automata.

\subsection{Determinization of functional group automata}

We first define a determinization construction for functional group
automata. This construction does not
necessarily yield a deterministic group automaton with a finite set of
states, unless, as we show in Subsection~\ref{sec:twinning}, the twinning property is satisfied.

The procedure is similar to the one of \cite{Droste_Kuich_Vogler_2009}
for weighted automata over a semiring (under some conditions, such as
weak divisibility), and to that of finite-state
transducers \cite{Choffrut77,BealEtAl03a}, but adapted to
groups. In particular, the procedure of \cite{Droste_Kuich_Vogler_2009} 
heavily uses the additive operation of semi-rings, and the
determinization of finite-state transducers relies on the operation of
taking the longest common prefix of two strings. One cannot rely on
similar operations for group automata. To be more precise, both known
determinization procedures in \cite{Choffrut77,BealEtAl03a,Droste_Kuich_Vogler_2009} extend the classical subset construction
with delays (what remains to be output). States are therefore pairs $(q,d)$ where
$q$ is a state of the original automaton, and $d$ is a delay (unlike
functional automata, there can be several delays associated with the
same state). Initially, all delays are equal to $\neutral$. 
When reading a new symbol $a$ from a state $P$, the
deterministic automaton outputs a value $v$ and goes to a state
$P'$. Let us explain how $v$ is computed. Let $T$ be the set of
transitions $(p,a,p')$ of the original automaton on $a$, such that
there exists $(p,d)\in P$. Let $\gamma(p,a,p')$ be their associated
value. The value $v$ is obtained by taking the sum (resp. the longest
common prefix) of the set $\{ d\cdot \gamma(p,a,p')\ |\ (p,d)\in
P\}$. In other words,  for each transition $(p,a,p')$, what remains to be output
is $d\cdot \gamma(p,a,p')$, and the output of the deterministic
automaton is somehow the ``best'' that can be output at the moment: the sum (or
longest common prefix respectively) of all these values. The state
$P'$ is then the set of states $p'$ with updated delays, i.e. what
still remain to be output.

Our determinization construction differs from  the procedures in \cite{Choffrut77,BealEtAl03a,Droste_Kuich_Vogler_2009} sketched above  in two points: $(i)$ it applies to
group automata that are \emph{functional}, and therefore there is at
most one delay per state (otherwise, since we assume that all state
are co-accessible, functionality would be falsified) $(ii)$ to compute
the value $v$, since we cannot rely on the additive operation, nor the 
longest common prefix operation, we pick a base transition
$(p_0,a,p'_0)$ (according to a given total order on transitions), and
$v$ is then $d\cdot \gamma(p_0,a,p'_0)$, where $d$ is the delay
associated with $p_0$. The new delays are computed relatively to $v$. 
The construction works for any order on transitions, possibly yielding
different deterministic automata\footnote{While point $(i)$ could be relaxed
by having sets of delays associated with each state, and several
outputs on the terminal symbol $\dashv$, we rather assume functionality 
as it is the purpose of this paper, and it lightens our notations. }.

Given the above premises, let us  describe our determinization construction formally. Let $A =
 (Q,q_I,F,\delta,\gamma)$ be a trim functional group automaton over a group
$(W,\cdot,\neutral)$ and assume, w.l.o.g.,  that  $\delta$ is totally ordered by some order
$\leq_\delta$. The output of our construction is  a group automaton
$A_d = (Q_d,f_I,F_d,\delta_d,\gamma_d)$ over $(W,\cdot,\neutral)$
such that $L_{A_d} = L_A$ and  $\delta_d$ is deterministic. However, $A_d$ is not a proper deterministic group automaton, as it may have infinitely many states.   The next section gives a sufficient condition under which 
$Q_d$ needs to be finite. 
Let $\mathcal{D}$  be the set of delays $\delay(\rho,\rho')$ for any
two runs $\rho,\rho'$ of $A$ on the same word. 
We  define $Q_d = \mathcal{D}^Q$, as the set of partial
functions from   states $Q$ to  delays.
%
%
%
%
%
%
We let $f_I : q_I\mapsto \neutral$ be the initial function, defined for
$q_I$ only. The set of accepting states $F_d$ is defined 
as $\{ f\in Q_d\ |\ \dom(f)\cap F\neq \varnothing\}$.
Then, given partial functions $f,f'\in Q_d$ and a symbol $a\in\Sigma$, 
we let $t_0 = (p'_0,a,p_0)$ be the smallest transition
(for $\leq_\delta$) from a state $p'\in \dom(f')$ to a state $p_0\in
\dom(f)$ on $a$. Let $v_0 =_{def} f'(p')\cdot
\gamma(t_0)$. Then, we let $(f',a,f)\in\delta_d$ if 
for all transitions $(q',a,q)\in\delta$ such that $q'\in \dom(f')$,
$f(q)$ is defined and equals
$$
f(q)=v_0^{-1}\cdot f'(q')\cdot  \gamma(q',a,q) 
$$
The value $\gamma_d(f',a,f)$ of the transition $(f',a,f)$ is
$v_0$. Note that $A_d$ is deterministic, because $f$ is functionally
obtained from $f'$ and $a$.

\begin{lem}\label{lem:eqAAd}
    Let $A$ be a trim functional group automaton and $A_d$ be the
    group automaton obtained by determinization. Then, for all $w\in \Sigma^+$, $L_A(w) = L_{A_d}(w)$. However, $A_d$ may
    have infinitely many states.
\end{lem}

\begin{proof}
    We show that $A_d$ maintains the following invariant: for all
    $w\in\Sigma^*$ such that there is a run $\rho_d : f_I\xrightarrow{w} f\in Q_d$, the 
    following three properties hold:
    \begin{enumerate}
        \item there exists $p_0\in \dom(f)$ such that
          $f(p_0)=\neutral$,
        \item for all $q\in\dom(f)$, for all runs
          $\rho:q_I\xrightarrow{w} q$ and $\rho_0 :
          q_I\xrightarrow{w} p_0$, $f(q) = \delay(\rho_0,\rho)$,
        \item for all $q\in\dom(f)$, for all runs
          $\rho:q_I\xrightarrow{w} q$,
          $V(\rho) = V(\rho_d)\cdot f(q)$. 
    \end{enumerate}

    We show this invariant by 
    induction on $|w|$. It is clearly true when $|w|=0$. 
    Suppose that $w = w'a$, where $a\in\Sigma$, and let $f'\in Q_d$ be
    such that $f_I\xrightarrow{w} f'$ and $(f',a,f)\in Q_d$. We show
    the two conditions:

    $(1)$ it suffices to take $p_0$ as defined in the determinization
    construction, because by definition $f(p_0) = v_0^{-1}\cdot v_0 =
    \neutral$.

    $(2)$ Let $\rho' : q_I\xrightarrow{w} q'$ be a run such that
    $\rho'.(q',a,q) = \rho$, and let $\rho'_0 : q_I\xrightarrow{w}
    p'$ be a run, where $p'$ is defined as in the definition of $\delta_d$. 
    By induction hypothesis, by $(1)$, there exists $p'_0\in Q$ such that 
    $f'(p'_0) = \neutral$. Let $r : q_I\xrightarrow{w} p'_0$ be a run of 
    $A$ on $w$. Again by induction hypothesis, by $(3)$, we have 
    $f'(p') = \delay(r, \rho'_0)$ and $f'(q') = \delay(r, \rho')$,
    i.e. $f'(p') = V(r)^{-1}\cdot V(\rho'_0)$ and $f'(q') =
    V(r)^{-1}\cdot V(\rho')$. Now, by definition of $\delta_d$, we
    have:
    $$
    \begin{array}{lllllllllll}
      f(q) & = & v_0^{-1}\cdot f'(q')\cdot \gamma(q',a,q) \\
      & = & (f'(p')\cdot \gamma(p',a,p_0))^{-1}\cdot f'(q')\cdot
            \gamma(q',a,q) \\
      & = & \gamma(p',a,p_0)^{-1}\cdot (f'(p'))^{-1}\cdot f'(q')\cdot
            \gamma(q',a,q)\\ 
      & = & \gamma(p',a,p_0)^{-1}\cdot (V(r)^{-1}\cdot V(\rho'_0))^{-1}\cdot V(r)^{-1}\cdot V(\rho')\cdot
            \gamma(q',a,q)\\ 
      & = & \gamma(p',a,p_0)^{-1}\cdot V(\rho'_0)^{-1}\cdot V(r)\cdot V(r)^{-1}\cdot V(\rho')\cdot
            \gamma(q',a,q)\\ 
      & = & \gamma(p',a,p_0)^{-1}\cdot V(\rho'_0)^{-1}\cdot V(\rho')\cdot
            \gamma(q',a,q)\\ 
           & = & V(\rho_0)^{-1}\cdot V(\rho) \\
           & = & \delay(\rho_0,\rho)
    \end{array}
    $$

    $(3)$ Let $\rho'_d$ be the run of $A_d$ on $w$. Let $\rho' :
    q_I\xrightarrow{w} q'$ be such that $\rho'.(q',a,q) = \rho$. By
    induction hypothesis and $(3)$, we know that 
    \begin{equation} \label{myeq1}
        V(\rho') = V(\rho'_d)\cdot f'(q')
    \end{equation}
    and therefore since $V(\rho_d) = V(\rho'_d)\cdot v_0$, 
    \begin{equation} \label{myeq2}
        V(\rho') = V(\rho_d)\cdot v_0^{-1}\cdot f'(q')
    \end{equation}
    From which we get
    \begin{equation} \label{myeq3}
        V(\rho')\cdot \gamma(q',a,q) = V(\rho_d)\cdot v_0^{-1}\cdot f'(q')\cdot \gamma(q',a,q)
    \end{equation}
    which, since $V(\rho) = V(\rho')\cdot \gamma(q',a,q)$ and 
    $f(q) = v_0^{-1}\cdot f'(q')\cdot \gamma(q',a,q)$, is equivalent
    to $V(\rho) = V(\rho_d)\cdot f(q)$.

    From this invariant, it is not difficult to show that $L_A =
    L_{A_d}$. Indeed, let $w\in \dom(A)$, let $\rho$ be an
    accepting run of $A$ on $w$ (therefore $V(\rho) = L_A(w)$), and
    let $\rho_d$ be the run of $A_d$ on $w$ (its existence is already
    a consequence of the subset construction in finite automata). 
    Clearly, $\rho_d$ ends in a state $f$ such that $\dom(f)\subseteq
    F$, because $w$ necessarily ends with $\dashv$, and it is the only
    way in $A$ to reach an accepting state. By invariant
    $(2)$, $f(q) = \delay(\rho_0,\rho)$ for $\rho_0$ a run of $A$ on
    $w$ ending in a state $p_0$ such that $f(p_0) = \neutral$. By
    invariant $(3)$, $V(\rho_d) = V(\rho_0)\cdot f(p_0) =
    V(\rho_0)$. Since $\dom(f)\subseteq F$, $\rho_0$ is accepting, and 
    since $A$ is functional, $V(\rho_0) = V(\rho)$. Therefore $V(\rho)
    = V(\rho_d)$, i.e. $L_{A_d}(w) = L_A(w)$. The converse is shown similarly.
\end{proof}

Note that for finite groups, this determinization always
yields an equivalent (finite) deterministic group automaton, because
the set of delays is finite. 

\begin{lem}
    Any functional group automaton over a finite group is
    determinizable. 
\end{lem}

\subsection{The twinning property: a sufficient condition for
  determinizability}\label{sec:twinning}

Lemma \ref{lem:finite} shows that the twinning property
(cfr. Definition \ref{twinningProp}) is sufficient to guarantee the 
termination of the determinization construction, i.e., the finiteness
of the automaton $A_d$ obtained by determinization. 

\begin{defi}\label{twinningProp}
Two states $p,q$ of a functional group automaton $A$ are \textit{twinned} if 
both $p$ and $q$ are co-accessible and for all words
$w_1,w_2\in\Sigma^*$, for all runs $\rho_1:q_I\xrightarrow{w_1}p$,
$\rho_2:p\xrightarrow{w_2} p$, $\rho'_1:q_I\xrightarrow{w_1}q$,
$\rho'_2:q\xrightarrow{w_2} q$, we have
$\delay(\rho_1,\rho'_1) = \delay(\rho_1\rho_2,\rho'_1\rho'_2)$.
The automaton $A$ satisfies the \textit{twinning property} if  all pairs of states are twinned.\end{defi}

\begin{lem}\label{lem:finite}
Let $A$ be a group automaton. If $A$ satisfies the twinning property,
then there are at most  $|\Sigma|^{|Q|^2}$ delays $\delay(\rho,\rho')$
for any two runs $\rho,\rho'$ on the same input, and thus the
deterministic group automaton $A_d$ obtained by the determinization procedure is finite.
\end{lem}
\begin{proof}
As delays must be equivalent on parallel loops, 
  any delay can be obtained with some pair of runs of length 
  $|Q|^2$ at most (on longer pairs of runs, there must exist a parallel
  loop with equivalent delays that can be removed without affecting the
  value of the global delay of both runs, see Lemma \ref{lem:removeslice}).
\end{proof}

The following lemma states a short witness property for the non
satisfiability of the twinning property. This result is crucial to
prove that the twinning property is decidable (cfr. Lemma
\ref{lem:decidetpds}).

\begin{lem}\label{lem:pumpingtp}
  If a functional group automaton $A$ does not satisfy the twinning
  property, there exist two words $w_1,w_2\in\Sigma^*$ such that
  $|w_1|\leq 2|Q|^2$ and $|w_2|\leq 2|Q|^2$, two states
  $p,q\in Q$ such that $p$ and $q$ are both co-accessible, and
  runs $\rho_1:q_I\xrightarrow{w_1}p$,
  $\rho_2:p\xrightarrow{w_2} p$, $\rho'_1:q_I\xrightarrow{w_1}q$,
  $\rho'_2:q\xrightarrow{w_2} q$, such that
  $\delay(\rho_1,\rho'_1) \neq \delay(\rho_1\rho_2,\rho'_1\rho'_2)$.
\end{lem}

\begin{proof}
  Suppose that $|w_2|>2|Q|^2$ (the case $|w_1|>2|Q|^2$ is 
  proved exactly the same way) and that $w_1w_2$ witnesses that the
  twinning property does not hold by the decomposition into runs
  $\rho_1,\rho_2,\rho'_1,\rho'_2$ as in the premisses of the lemma. 
  We will show that we can shorten the runs $\rho_1,\rho'_1$ and still
  get a witness that the twinning property does not hold.

  Since $|w_2|>2|Q|^2$, there is a pair of states $(p',q')$ that
  repeats three times along the two parallel runs $\rho_2$ and
  $\rho'_2$, i.e. $w_2$ can be decomposed as
  $w'_1w'_2w'_3w'_4$ and $\rho_2$ and $\rho'_2$ can
  be decomposed as $r_1r_2r_3r_4$ and $r'_1r'_2r'_3r'_4$ respectively, where:
  $$
  \begin{array}{lllllll}
  r_1: p\xrightarrow{w'_1} p' & r_2 : p'\xrightarrow{w'_2} p' & r_3:
  p'\xrightarrow{w'_3} p' & r_4 : p'\xrightarrow{w'_4} p \\
  r'_1: q\xrightarrow{w'_1} q' & r'_2 : q'\xrightarrow{w'_2} q' & r'_3:
  q'\xrightarrow{w'_3} q' & r'_4 : p'\xrightarrow{w'_4} q 
  \end{array}
  $$

  Note that $r_1,r'_1$ and $r_4,r'_4$ may be empty (in this case $p = p'$ and
  $q=q'$), but $r_2,r_3,r'_2,r'_3$ are assumed to be non-empty.

  Now, there are two cases: $\delay(\rho_1r_1,\rho'_1r'_1)\neq \delay(\rho_1r_1r_2,\rho'_1r'_1r'_2)$ and 
  in that case the word $w_1w'_1w'_2$ is a witness that the twinning
  property does not hold, and $|w_1w'_1w'_2| < |w_1w_2|$. 
  In the second case, we have
  $\delay(\rho_1r_1,\rho'_1r'_1) =\delay(\rho_1r_1r_2,\rho'_1r'_1r'_2)$, but in that case, we can
  apply Lemma \ref{lem:removeslice} and we get
  $\delay(\rho_1r_1r_3r_4,\rho'_1r'_1r'_3r'_4) =
  \delay(\rho_1\rho_2,\rho'_1\rho'_2)$. 
  
  Therefore, 
$\delay(\rho_1,\rho'_1)\neq
  \delay(\rho_1r_1r_3r_4,\rho'_1r'_1r'_3r'_3)$ and $w_1w'_1w'_3w'_4$
  is a shorter witness that the twinning property does not hold.
\end{proof}

\begin{lem}\label{lem:decidetpds}
  It is decidable in {\sf CoNP} whether a functional group
  automaton satisfies the
  twinning property.
\end{lem}
\begin{proof}
First, the automaton is transformed into a trim automaton.
Then, it suffices to guess two runs 
on the same input word of size at most $4|Q|^2$ and two positions in
those runs, and check (in ptime) that the pair of states at the two positions are
equal and that the respective delays are different. This algorithm
is correct thanks to Lemma \ref{lem:pumpingtp}.
\end{proof}

\subsection{Determinization of functional infinitary group automata}

In this subsection, we define the class of infinitary groups, and show
that the twinning property is a necessary condition for
determinizability of functional group automata over an infinitary
group. Morevoer, we show that the groups encoding 
\Sum, \Avg, and \Dsum (cf. Lemma \ref{lem:encodegroup}) belong
to this class, yielding a procedure to decide determinizability for
the classes of weighted $V$-automata, for $V\in\{\Sum,\Avg,\Dsum\}$. 
Intuitively, the infinitary condition implies that iterating two runs
of an infinitary group automaton  on a parallel loop induces
infinitely many delays.

\begin{defi}[Infinitary Groups]\label{def:infinitary}
A group $(W,\cdot,\neutral)$ is said to be \emph{infinitary} if it satisfies
the \emph{infinitary condition} stating that for all
$v_1,w_1,v_2,w_2\in W$, if $v_1^{-1}\cdot w_1\neq v_2^{-1}\cdot
v_1^{-1}\cdot w_1\cdot w_2$, then:
$$|\{v_2^{-h}\cdot v_1^{-1}\cdot w_1\cdot w_2^h\:|\: h\geq 0\}|=\infty$$
A group automaton over an infinitary group is called an \emph{infinitary
group automaton}. 
\end{defi}

We show that the groups encoding \Sum, \Avg and \Dsum satisfy the
infinitary condition, as well as any linearly ordered group. 
A  \emph{linearly ordered group} $(W,\cdot,\neutral,<)$ is simply a group
  $(W,\cdot,1)$ equipped with a strict total order $<$ such that
  (monotonicity) for all $ x,y,l,r\in W$, if $x< y$ then $l\cdot
x\cdot r < l\cdot y\cdot r$. 

\begin{prop}\label{prop:linear}
Any linearly ordered group $(W,\cdot,\neutral,<)$ is
    infinitary.
\end{prop}

\begin{proof}
 Indeed, let $v_1,w_1,v_2,w_2\in W$ such that
    $v_1^{-1}\cdot w_1\neq v_2^{-1}\cdot v_1^{-1}\cdot w_1\cdot
    w_2$. Suppose that $v_1^{-1}\cdot w_1< v_2^{-1}\cdot
    v_1^{-1}\cdot w_1\cdot w_2$, then by monotonicity, 
    $v_2^{-1}\cdot v_1^{-1}\cdot w_1 \cdot w_2<  v_2^{-1}\cdot
    v_2^{-1}\cdot v_1^{-1}\cdot w_1\cdot w_2\cdot w_2$. More
    generally, for all $i<j$, $v_2^{-i}\cdot v_1^{-1}\cdot w_1\cdot
    w_2^i < v_2^{-j}\cdot v_1^{-1}\cdot w_1\cdot w_2^j$. The
    infinitary condition is shown similarly when assuming 
    $v_1^{-1}\cdot w_1> v_2^{-1}\cdot v_1^{-1}\cdot w_1\cdot w_2$.
\end{proof}

Since the groups encoding \Sum and \Avg can be linearly ordered, we
immediately get that they are infinitary. For \Dsum, we do a different
proof. 

\begin{prop}
    For $V\in\{\Sum,\Avg,\Dsum\}$, let $G_V$ be the groups defined 
    in Lemma \ref{lem:encodegroup}. Then $G_V$ is infinitary.
\end{prop}

\begin{proof}
    We show that for the three measures, the groups that encode them
    (see Lemma \ref{lem:encodegroup}) can be linearly ordered. Then,
    the result follows by Proposition \ref{prop:linear}. 

    For $G_\Sum$, it suffices to order its elements with the natural
    order on integers. Similarly, for $G_\Avg$, it suffices to order its elements
    by lexicographic order. For both orders, it is easy to show they
    satisfy the monotonicity property.

    The group $G_\Dsum$ can also be linearly
    ordered by lexicographic order as follows: $(a,x)<(b,y)$ if $x<y$,
    or $x=y$ and $a<b$. Let us show that this order satisfies the
    monotonicity condition. Assume that $(a,x)<(b,y)$ and 
    consider $(c,z)\in
    \mathbb{Q}\times\mathbb{Q}^+$. Then, $(c,z)\cdot (a,x) =
    (\frac{c}{x}+a, zx)$ and 
    $(c,z)\cdot (b,y) =
    (\frac{c}{y}+b, zy)$. Suppose that $x<y$, then clearly, since
    $z,x,y$ are strictly positive, we have $zx<zy$, and therefore 
    $(c,z)\cdot (a,x) < (c,z)\cdot (b,y)$. Otherwise, if $x=y$, then
    $a<b$, and we get $zx = zy$, $\frac{c}{x} + a < \frac{c}{y} + b$,
   which implies again $(c,z)\cdot (a,x) < (c,z)\cdot (b,y)$. 

   Now, we show that $(a,x)\cdot (c,z) < (b,y) \cdot  (c,z)$. By
   definition of $\cdot$, $(a,x)\cdot (c,z) = (\frac{a}{z}+c, xz)$ and 
   $(b,y)\cdot (c,z) = (\frac{b}{z}+c, yz)$. If $x<y$, then clearly
   $xz<yz$ and we are done. Otherwise, $x=y$ and $a<b$, which implies
   that $\frac{a}{z}+c < \frac{b}{z}+c$.
\end{proof}

\begin{lem}\label{lem:notdet}
Let $A$ be a trim functional infinitary group automaton.
If $A$ does not satisfy the twinning property, then $A$ is not determinizable.
\end{lem}
\begin{proof}
   Suppose that the twinning property does not hold. Therefore 
   we can find words $w_1$, $w_2$  and runs $\rho_1,\rho'_1$ on
   $w_1$ and $\rho_2,\rho'_2$ on $w_2$ such that
   $\rho_2, \rho'_2$ are loops (on state $p$ and $q$ respectively), and
   $\delay(\rho_1,\rho'_1)\neq
   \delay(\rho_1\rho_2,\rho'_1\rho'_2)$. Let
   $\Delta(k) = \delay(\rho_1(\rho_2)^k, \rho'_1(\rho'_2)^k)$ for all
   $k\in\mathbb{N}$. By the infinitary condition, the set 
   $\{\Delta(k)\ |\ k\geq 0\}$ is infinite. Let $S\subseteq
   \mathbb{N}$ such that for all $i,j\in S$ such that $i\neq j$, 
   $\Delta(i)\neq \Delta(j)$. Note that $S$ is infinite.

   Suppose that there exists a deterministic
   automaton $A_d = (Q_d,f_I,F_d,\delta_d,\gamma_d)$ equivalent to
   $A$. We exhibit a contradiction. Since $S$ is infinite and $Q_d$ is
   finite, we can find $k_1\neq k_2\in \mathbb{N}$ such that $k_1\in
   S$, $k_1+k_2\in S$ (i.e. $\Delta(k_1)\neq \Delta(k_1+k_2)$), and such
   that on $w_1(w_2)^{k_1+k_2}$, the run of $A_d$ has the following form:
   $$
  f_I\xrightarrow{w_1(w_2)^{k_1}}
  f\xrightarrow{(w_2)^{k_2}} f
  $$
  for some $f\in Q_d$.

  Moreover, since $p$ and $q$ are both co-accessible, there exist
  two words $w_3,w'_3$ and two runs of $A_d$ of the form:
  $$
    \begin{array}{lllllll}
      \rho_d\ :\ f_I\xrightarrow{w_1(w_2)^{k_1}}
      f\xrightarrow{(w_2)^{k_2}}
      f\xrightarrow{w_3} g \\
      \rho'_d\ :\ f_I\xrightarrow{w_1(w_2)^{k_1}}
      f\xrightarrow{(w_2)^{k_2}}
      f\xrightarrow{w'_3} g' \\
    \end{array}
  $$
    for some accepting states $g,g'\in F_d$. Let
    $\rho_d = \rho_{d,1}\rho_{d,2}\rho_{d,3}$ and $\rho'_d =
    \rho_{d,1}\rho_{d,2}\rho'_{d,3}$ for some subruns
    $\rho_{d,1},\rho_{d,2}$ that correspond to $w_1(w_2)^{k_1}$ and
    $(w_2)^{k_2}$ respectively, and some subruns $\rho_{d,3}$ and
    $\rho'_{d,3}$ that correspond to $w_3$ and 
    $w'_3$ respectively.
    We know that $\Delta(k_1)\neq \Delta(k_1+k_2)$.
    We show that this leads to a contradiction. Let also
    $\rho_3:p\xrightarrow{w_3} p_f$ and $\rho'_3:q\xrightarrow{w'_3}
    q_f$ be two runs in $A$, for some $p_f,q_f\in F$. Then we have:
    \begin{equation}\label{eq:z1}
      V(\rho_{d,1}\rho_{d,2}\rho_{d,3})\ =\ V(\rho_1(\rho_2)^{k_1+k_2}\rho_3)
    \end{equation}
    \begin{equation}\label{eq:z2}
     V(\rho_{d,1}\rho_{d,2}\rho'_{d,3})\ =\ V(\rho'_1(\rho'_2)^{k_1+k_2}\rho'_3)
    \end{equation}
    \begin{equation}\label{eq:z3}
      V(\rho_{d,1}\rho_{d,3})\ =\ V(\rho_1(\rho_2)^{k_1}\rho_3)
    \end{equation}
    \begin{equation}\label{eq:z4}
      V(\rho_{d,1}\rho'_{d,3})\ =\ V(\rho'_1(\rho'_2)^{k_1}\rho'_3)
    \end{equation}
   From Equations \ref{eq:z1} and \ref{eq:z2} we get:
    \begin{equation}
        V(\rho_{d,1}\rho_{d,2}\rho'_{d,3})^{-1}\cdot V(\rho_{d,1}\rho_{d,2}\rho_{d,3})
    = V(\rho'_1(\rho'_2)^{k_1+k_2}\rho'_3)^{-1}\cdot V(\rho_1(\rho_2)^{k_1+k_2}\rho_3)
    \end{equation}
    which is equivalent to:
    \begin{equation}\label{eq:delta4}
        V(\rho'_{d,3})^{-1}\cdot V(\rho_{d,3})
    = V(\rho'_1(\rho'_2)^{k_1+k_2}\rho'_3)^{-1}\cdot V(\rho_1(\rho_2)^{k_1+k_2}\rho_3)
    \end{equation}
    Similarly from Equations \ref{eq:z3} and \ref{eq:z4} we get:
    \begin{equation}\label{eq:delta5}
        V(\rho'_{d,3})^{-1}\cdot V(\rho_{d,3})
    = V(\rho'_1(\rho'_2)^{k_1}\rho'_3)^{-1}\cdot V(\rho_1(\rho_2)^{k_1}\rho_3)
    \end{equation}
    From Equations \ref{eq:delta4} and \ref{eq:delta5} we get:
    \begin{equation}\label{eq:delta6}
V(\rho'_1(\rho'_2)^{k_1+k_2}\rho'_3)^{-1}\cdot V(\rho_1(\rho_2)^{k_1+k_2}\rho_3) =V(\rho'_1(\rho'_2)^{k_1}\rho'_3)^{-1}\cdot V(\rho_1(\rho_2)^{k_1}\rho_3)
    \end{equation}
    By simplifying this expression we obtain:
    \begin{equation}\label{eq:delta7}
V((\rho'_2)^{k_2})^{-1}\cdot V(\rho'_1)^{-1}\cdot V(\rho_1)\cdot V((\rho_2)^{k_2}) =V(\rho'_1)^{-1}\cdot V(\rho_1)
    \end{equation}
    And therefore by multiplying by $V((\rho'_2)^{k_1})^{-1}$ to the left and by 
    $V((\rho_2)^{k_1})$ to the right, we obtain:
    \begin{equation}\label{eq:delta8}
    \begin{array}{c}
V((\rho'_2)^{k_1+k_2})^{-1}\cdot V(\rho'_1)^{-1}\cdot V(\rho_1)\cdot V((\rho_2)^{k_1+k_2}) 
= \\ =V((\rho'_2)^{k_1})^{-1}\cdot V(\rho'_1)^{-1}\cdot V(\rho_1)\cdot V((\rho_2)^{k_1})
\end{array}
    \end{equation}
    which can be rewritten as:
    \begin{equation}\label{eq:delta9}
V(\rho'_1(\rho'_2)^{k_1+k_2})^{-1}\cdot V(\rho_1(\rho_2)^{k_1+k_2}) =V(\rho'_1(\rho'_2)^{k_1})^{-1}\cdot V(\rho_1(\rho_2)^{k_1})
    \end{equation}
    which is equivalent to $\Delta(k_1) = \Delta(k_1+k_2)$. This contradicts our hypothesis.
\end{proof}

Therefore, we obtain:

\begin{thm}\label{thm:detdsum}
    A functional infinitary group automaton is (effectively) determinizable iff it
    satisfies the twinning property. Determinizability is
    decidable in {\sf CoNP} for functional infinitary group automata,
    and functional $V$-automata, for $V\in\{\Sum,\Avg,\Dsum\}$.
\end{thm}
\begin{proof}
It follows directly from Lemmas \ref{lem:finite}-\ref{lem:notdet} and
Lemma \ref{lem:encodegroup}.
\end{proof}


\section{Conclusion}
\newcommand{\rotateangle}{300}
\newcommand{\spacetable}{1mm}

\begin{table}[t]
{\centering
        \begin{tabular}{l||c@{\hspace{\spacetable}}c@{\hspace{\spacetable}}c@{\hspace{\spacetable}}|c@{\hspace{\spacetable}}c@{\hspace{\spacetable}}c@{\hspace{\spacetable}}|c@{\hspace{\spacetable}}c@{\hspace{\spacetable}}c@{\hspace{\spacetable}}}

         Decision Problems & \multicolumn{3}{c}{\Sum/\Avg-Automata} 
         & \multicolumn{3}{c}{\Dsum-Automata} &
         \multicolumn{3}{c}{\Ratio-Automata} \\
          & 
         \textsf{D} & \textsf{F} & \textsf{ND} & 
               \textsf{D} & \textsf{F} & \textsf{ND} & 
          \textsf{D} & \textsf{F} & \textsf{ND} \\

        \hline functionality &
        - & - & \textsf{P} & 
        - & - & \textsf{P} & 
        - & - & \textsf{CoNP}
        \\

        ${>}\nu$-emptiness & 
        \textsf{P} & \textsf{P}  & \textsf{P}\cite{ChatterjeeDH10} &
        \textsf{P} & \textsf{P}  & \textsf{P} &
        \textsf{P} & \textsf{P}  & \textsf{P} 
        \\

        ${\geq}\nu$-emptiness & 
        \textsf{P} & \textsf{P}  & \textsf{P}\cite{ChatterjeeDH10} &
        {?} & \textsf{PSpace}\cite{boker1}  & {?} &
        \textsf{P} & \textsf{P}  & \textsf{P} 
        \\

        ${>}\nu$-universality & 
        \textsf{P} & \textsf{P}  & \textsf{U}\cite{Krob/94,AlmagorBokerKupferman2011}  &
        {?} & \textsf{PSpace}\cite{boker1}  & {?} &
        \textsf{P} & \textsf{P}  & \textsf{P} 
        \\

        ${\geq}\nu$-universality & 
        \textsf{P} & \textsf{P}  & \textsf{U}\cite{Krob/94,AlmagorBokerKupferman2011}  &
        \textsf{P} & \textsf{P}  & \textsf{P} &
        \textsf{P} & \textsf{P}  & \textsf{P} 
        \\

        inclusion $A\leq B$ &
        \textsf{P} & \textsf{PSpace-c}  & \textsf{U}\cite{Krob/94} & 
        \textsf{P} & \textsf{PSpace-c}  & {?} & 
        \textsf{D} & \textsf{D}  & \textsf{U} 
        \\
        ($A$ arbitrary) & & & & & & & & & \\

        equivalence & 
        \textsf{P} & \textsf{PSpace-c}  & \textsf{U}\cite{Krob/94} & 
        \textsf{P} & \textsf{PSpace-c}  & {?} & 
        \textsf{D} & \textsf{D}  & \textsf{U} 
        \\

        determinizability &
        - & \textsf{CoNP}  & {?} & 
        - & \textsf{CoNP}  & {?} &
        - & {?}  & {?}
        \\

        ${>}\nu$-realizability & 
        \textsf{NP}$\cap$ & \textsf{U}  & \textsf{U} &
        \textsf{NP}$\cap$ & {?}  & {?} &
        \textsf{NP}$\cap$ & \textsf{U}  & \textsf{U} \\

          & \textsf{CoNP} & & & 
\textsf{CoNP} & & & 
\textsf{CoNP} & & \\
         \bottomrule
        \end{tabular}   
\caption{\label{tbl:summary} Complexity results for classes of
  weighted automata. In this table, \textsf{D} stands for
  deterministic, \textsf{F} for functional and \textsf{ND} for
  non-deterministic. The symbol \textsf{P} stands \textsf{PTime}, \textsf{U} for undecidable, \textsf{D}
  for decidable and ? for open.}}
\end{table}

In this paper, we have introduced and studied classes of
\emph{functional} weighted automata, for four measures: 
sum, average, discounted sum and ratio. Our results are summarized in
Table~\ref{tbl:summary}. Despite our efforts, some problems remain
open. 

The determinizability of (non-deterministic) \Sum-automata is a
long standing open problem \cite{journals/mst/Kirsten06,AlmagorBokerKupferman2011}.  Another notorious open problem,  the so-called target discounted-sum problem \cite{boker1}, underlies a number of open questions related to  \Dsum automata and games: namely, e.g.  imperfect information   \Dsum-games \cite{DDGRT10}, multi-objective \Dsum-games \cite{Chatterjee13}, universality of \Dsum-automata \cite{DBLP:conf/csl/BokerH11}, as well as  the realizability problem for \Dsum-automata that we mention here. Another open problem, that we
did not mention so far, is to decide whether a (non-deterministic) \Sum-automaton is
equivalent to some functional \Sum-automaton. Solving this problem would
solve the determinizability problem for non-deterministic 
\Sum-automata, as the determinizability problem for functional
\Sum-automata is decidable. However, this problem does not seem to be
simpler than the determinizability problem. Finally, we leave 
the determinizability problem for functional \Ratio-automata as
open: the techniques we developed in this paper  toward determinization are based on the
well-known  notion of delay between runs. It is not clear what would be a
suitable notion of delays for \Ratio-automata.

%


\section*{Acknowledgement}
We are very grateful to the anonymous
referees for their valuable comments on preliminary versions of this
paper. 

\bibliographystyle{plain}

\begin{thebibliography}{10}

\bibitem{AlmagorBokerKupferman2011}
S. Almagor, U. Boker, and O. Kupferman.
\newblock What's decidable about weighted automata?
\newblock In ATVA'11, pp 482--491.
  Springer, 2011.

\bibitem{AminofKL11}
B. Aminof, O. Kupferman, and R. Lampert.
\newblock Rigorous approximated determinization of weighted automata.
\newblock In {\em LICS}, pp 345--354, 2011.

\bibitem{andersson2006}
D. Andersson.
\newblock An improved algorithm for discounted payoff games.
\newblock In {\em ESSLLI Student Session}, pp 91--98, 2006.

\bibitem{BealEtAl03a}
M.-P. Beal, O. Carton, C. Prieur, and J. Sakarovitch.
\newblock Squaring transducers: An efficient procedure for deciding
  functionality and sequentiality.
\newblock {\em TCS}, 292, 2003.

\bibitem{BerstelReutenauer88}
J. Berstel and C. Reutenauer.
\newblock {\em Rational Series and Their Languages}.
\newblock Number~12 in EATCS Monographs on TCS.
  Springer, 1988.

\bibitem{JCSS::BlattnerH1977}
M. Blattner and T. Head.
\newblock Single-valued {$a$}-transducers.
\newblock {\em JCSS}, 15(3), pp 310--327, 1977.

\bibitem{BloemGHJ09}
R. Bloem, K. Greimel, T. A. Henzinger, and B. Jobstmann.
\newblock Synthesizing robust systems.
\newblock In {\em FMCAD}, pp 85--92. IEEE, 2009.

\bibitem{BokerCHK11}
U. Boker, K. Chatterjee, T.~A. Henzinger, and O. Kupferman.
\newblock Temporal specifications with accumulative values.
\newblock In {\em LICS}, pp 43--52, 2011.

 \bibitem{boker1}
U. Boker, T. Henzinger and J. Otop.
\newblock The Target Discount-Sum Problem.
\newblock In {\em LICS}. To appear, 2015.

\bibitem{DBLP:conf/csl/BokerH11}
U. Boker and T.~A. Henzinger.
\newblock Determinizing discounted-sum automata.
\newblock In {\em CSL}, pp 82--96, 2011.


\bibitem{ChatterjeeDH10}
K. Chatterjee, L. Doyen, and T.~A. Henzinger.
\newblock Quantitative languages.
\newblock {\em ACM Trans. Comput. Log}, 11(4), 2010.


\bibitem{Chatterjee13}
K. Chatterjee, V. Forejt, and D. Wojtczak.
\newblock Multi-objective Discounted Reward Verification in Graphs and MDPs.
\newblock {\em Logic for Programming, Artificial Intelligence, and
  Reasoning}, LNCS 8312, 2013.


\bibitem{Choffrut77}
C. Choffrut.
\newblock Une caract\'erisation des fonctions s\'equentielles et des fonctions
  sous-s\'equentielles en tant que relations rationnelles.
\newblock {\em TCS}, 5(3), pp 325--337, 1977.

\bibitem{AlfaroFHMS05}
L. de~Alfaro, M. Faella, T.~A. Henzinger, R. Majumdar, and
  M. Stoelinga.
\newblock Model checking discounted temporal properties.
\newblock {\em TCS}, 345(1), pp 139--170, 2005.

\bibitem{DDGRT10}
A. Degorre, L. Doyen, R. Gentilini, J-F. Raskin and S. Torunczyk
\newblock Energy and Mean-Payoff Games with Imperfect Information.
\newblock In {\em CSL-2010}, LNCS vol. 6247, pp 260-274, 2010.


\bibitem{Droste_Kuich_Vogler_2009}
M. Droste, W. Kuich, and H. Vogler.
\newblock {\em Handbook of Weighted Automata}.
\newblock Springer, 2009.

\bibitem{DBLP:conf/wia/DrosteR07}
M. Droste and G. Rahonis.
\newblock Weighted automata and weighted logics with discounting.
\newblock In {\em CIAA}, pp 73--84, 2007.

\bibitem{EM79}
A. Ehrenfeucht and J. Mycielski. 
\newblock Positional strategies for mean payoff games. 
\newblock In {\em International Journal of Game Theory}, 8(2):109–113, 1979.

\bibitem{FGR12}
E. Filiot, R. Gentilini and  J-F. Raskin
\newblock Quantitative languages defined by functional automata.
\newblock In {\em CONCUR-2012}, LNCS vol. 7454, pp 132-146, 2012.


\bibitem{Grunewald04}
F. Grunewald and D. Segal.
\newblock On the integer solutions of quadratic equations.
\newblock {\em Journal f\"ur die reine und angewandte Mathematik}, 569, pp 13--45,
  2004.

\bibitem{journals/mst/GurariI83}
E.~Gurari and O.~Ibarra.
\newblock A note on finitely-valued and finitely ambiguous transducers.
\newblock {\em Mathematical Systems Theory}, 16(1):61--66, 1983.

\bibitem{journals/mst/Kirsten06}
D. Kirsten.
\newblock A burnside approach to the termination of mohri's algorithm for
  polynomially ambiguous min-plus-automata.
\newblock {\em ITA}, 42(3), pp 553--581, 2008.

\bibitem{KirstenM05}
D. Kirsten and I. M{\"a}urer.
\newblock On the determinization of weighted automata.
\newblock {\em Journal of Automata, Languages and Combinatorics},
  10(2/3), pp 287--312, 2005.

\bibitem{DBLP:journals/tcs/KlimannLMP04}
I. Klimann, S. Lombardy, J. Mairesse, and C. Prieur.
\newblock Deciding unambiguity and sequentiality from a finitely ambiguous
  max-plus automaton.
\newblock {\em TCS}, 327(3), pp 349--373, 2004.

\bibitem{Krob/94}
D. Krob.
\newblock The equality problem for rational series with multiplicities in the
  tropical semiring is undecidable.
\newblock {\em Journal of Algebra and Computation}, 4(3), pp 405--425, 1994.

\bibitem{Krob94someconsequences}
D. Krob and P. Litp.
\newblock Some consequences of a fatou property of the tropical semiring.
\newblock {\em J. Pure Appl. Algebra}, 93, pp 231--249, 1994.

\bibitem{DBLP:journals/ita/LombardyM06}
S. Lombardy and J. Mairesse.
\newblock Series which are both max-plus and min-plus rational are unambiguous.
\newblock {\em ITA}, 40(1), pp 1--14, 2006.

\bibitem{minsky67finite}
M.~L. Minsky.
\newblock {\em Computation: Finite and Infinite Machines}.
\newblock Prentice-Hall, 1967.

 \bibitem{Mohri97}
 M. Mohri.
 \newblock Finite-State Transducers in Language and Speech Processing.
 \newblock {\em Computational Linguistics}, pp 269--311, 1997.

 \bibitem{Mohri09}
 M. Mohri.
 \newblock Weighted automata algorithms.
 \newblock {\em Handbook of Weighted Automata}, pp 213--254, 2009.

 \bibitem{Parikh}
R. J. Parikh
\newblock On context-free languages.
\newblock {\em Journal of the ACM}, 13 (4) (1966) 570–581.

\bibitem{PnuRos:89}
A.~Pnueli and R.~Rosner.
\newblock On the synthesis of a reactive module.
\newblock In {\em ACM Symposium on Principles of Programming Languages (POPL)}.
  ACM, 1989.

\bibitem{Schmidt77}
E.~M. Schmidt and T.~G. Szymanski.
\newblock Succinctness of descriptions of unambiguous context-free
languages.
\newblock In {\em SIAM Journal on Computing}, 6(3), pp 547--553, 1977.


\bibitem{Schutz75}
M.~P. Sch{\"u}tzenberger.
\newblock Sur les relations rationnelles.
\newblock In {\em Automata Theory and Formal Languages}, vol~33 of {\em
  LNCS}, pp 209--213, 1975.

\bibitem{DBLP:conf/birthday/Thomas08}
W. Thomas.
\newblock Church's problem and a tour through automata theory.
\newblock In {\em Pillars of Computer Science}, pp 635--655, 2008.

\bibitem{ACTAI::Weber1989}
A.~Weber.
\newblock On the valuedness of finite transducers.
\newblock {\em Acta Informatica}, 27(8):749--780, 1989.

\bibitem{DBLP:journals/iandc/WeberK95}
A. Weber and R. Klemm.
\newblock Economy of description for single-valued transducers.
\newblock {\em Inf. Comput.}, 118(2), pp 327--340, 1995.

\bibitem{KariantoKT06}
K. Wong, A. Krieg, and W. Thomas.
\newblock On intersection problems for polynomially generated sets.
\newblock In {\em ICALP}, vol 4052 of {\em LNCS}, pp 516--527. Springer,
  2006.
 
 \bibitem{ZP96} 
 U. Zwick and M. Paterson. 
\newblock  The complexity of mean payoff games on graphs. 
\newblock In {\em Theoretical Computer Science}, 158(1-2):343–359, 1996.

\end{thebibliography}

\end{document}